\documentclass[onecolumn,times, a4paper]{article}

\usepackage [a4paper, total={210mm,397mm}, left=30mm, right=30mm, top=20mm, bottom=25mm]{geometry}
\usepackage{amsmath}
\usepackage{amsthm}
\usepackage{amssymb,amsfonts,amsmath,amsthm, amsbsy}
\usepackage{graphicx}
\usepackage{colortbl}
\usepackage{color}
\usepackage[most]{tcolorbox}
\graphicspath{ {./} }
\usepackage[normalem]{ulem}

\usepackage{listings}
\usepackage{float}
\usepackage{xspace}
\lstdefinestyle{mystyle}{
	backgroundcolor=\color{gray!5},
	numbers=left,
	frame=tb,
	keywordstyle=\color{black}\bfseries\normalsize,
	basicstyle=\sffamily\small\color{black},
	commentstyle=\color{gray},
	breaklines=true,
	numberstyle=\bfseries\footnotesize\color{gray}
}

\newtheorem {Def} {Definition}
\newtheorem {Teo} {Theorem}
\newtheorem {Lem} {Lemma}
\newtheorem {Prop}{Property}
\newtheorem {Note}{Note}

\newtheorem {Con} {Conjecture}

\DeclareMathOperator*{\argmax}{arg\,max}

\def\path     {{path}}

\def\paths    {{paths}}
\def\track    {{path}}
\def\tracks   {{paths}}
\def\subtrack {{subpath}}

\def\subpaths {{subpaths}}
\def\subpath  {{subpath}}
\def\concat   {{concatenation}}
\def\concats  {{concatenations}}

\def\lpath {{\sf l_p}}
\def\rpath {{\sf r_p}}
\def\lPmax {\textit {lPmax}}
\def\rPmax {\textit {rPmax}}

\def\O     {{\textit O}}
\def\H     {{\textit H}}
\def\W     {{\textit W}}
\def\PS    {{\sf PS}}
\def\F     {{\sf F}}
\def\g     {{\sf g}}
\def\lg     {{\sf g_l}}
\def\rg     {{\sf g_r}}
\def\lrtms {{\sf lrdtms}}
\def\LP    {{\sf LP}}
\def\RP    {{\sf RP}}

\def\CP    {{CP2}}

\def\lMaxes{{\sf {M_l}}}
\def\rMaxes{{\sf {M_r}}}

\usepackage{csquotes}
\title{
On Linear Solution of \textquote{Cherry Pickup II}.\\
Max Weight of Two Disjoint Paths\\
in Node-Weighted Gridlike DAG
}
\author{Igor N. Tunev}
\date{\today}
\date{itnvi@mail.ru}

\begin{document}
\maketitle

\begin{abstract}
\textquote{Minimum Falling Path Sum} (MFPS) is classic question in programming --
\textquote{Given a grid of size $N{\times}N$ with integers in cells, return the minimum sum of a falling path through grid.
A falling path starts at any cell in the first row and ends in last row, with the rule of motion -- the next element after the cell $(i, j)$ is one of the cells $(i + 1, j - 1)$, $(i + 1, j)$ and $(i + 1, j + 1)$}.
This problem has linear solution (LS) (i.e. $\O(N^2)$) using dynamic programming method (DPM).

There is an Multi-Agent version of MFPS called \textquote{Cherry Pickup II} (\CP)~\cite{bib:CPII}.
\CP\ is a search for the maximum sum of 2 falling paths started
from top corners, where each covered cell summed up one time.
All known fast solutions of \CP\ uses DPM, but have $\O(N^3)$ time
complexity on grid ${N\times}N$.
Here we offer a LS of \CP\ (also using DPM) as finding maximum total weight of 2 vertex-disjoint paths.
Also, we extend this LS for some extended version of \CP\ with wider motion rules.
\end{abstract}


\section{Introduction}

\CP\ is Multi-Agent extension of well known problem, sometimes called as \textquote{Minimum Falling Path Sum}
in ~\cite{bib:MFPS},
and its variations like \textquote{Gold Mine} in~\cite{bib:gold_mine} and \textquote{Minimum Path Sum} in~\cite{bib:MPS}.

%Original description of \CP\ can be found in~\cite{bib:CPII}.
There is variation of \CP\ called \textquote{Cherry Pickup} in~\cite{bib:CP_leetcode}
sometimes called as \textquote{Diamond Mine} (DM) in~\cite{bib:DM}.
DM extended with ability to lock cells, but still has linear reducing to \CP,
even as finding maximum sum of 2 node-disjoint paths, as will be described below.

For solution of \CP\ we offer algorithm for search of 2 paths without crossing with maximum common sum.
Thus, this LS can be represented as LS for a simple case of Multi-Agent Path Finding
problem (MAPF) with maximizing/minimizing deliveries/cost.
The MAPF is the problem of finding collision-free paths for a team of robots from
their locations to given destinations in a known environment.

Disjoint paths (DP) problem is one of the well known problems in algorithmic graph theory and combinatorial optimization.
There are many LSs of finding fixed number of DP on spetial cases of graphs.
For example, Scheffler found LS on graphs with bounded tree-width~\cite{bib:DP_BTW}.
In the paper of Golovach, Kolliopoulos, Stamoulis and Thilikos~\cite{bib:DP_PG} offered
LS on a planar graphs.
Most closely for our purpose is LS proposed by Tholey for 2 DP on directed acyclic graphs (DAGs) ~\cite{bib:DP_DAG}.
But we need in LS on node- or edge-weighted DAGs.

Suitable for our purpose the Suurballe's algorithm (SA)
on edge-weighted digraphs ~\cite{bib:SUURB},
but with not linear complexity, as we will show further.
We offer LS for finding 2 node-DP with maximum total weight on some special case of node-weighted DAGs.

\subsection{Problem description}
Given a grid $\g$ of size $\H{\times}\W$ with addressable cells from $(0,0)$ to $(\H{-}1,\W{-}1)$.
Each cell in grid represents the number of cherries that we can collect.
There are 2 robots in corners $(0,0)$ and $(0,\W{-}1)$, that can collect cherries.
When a robot is located in a cell, It picks up all cherries of this cell,
and this cell becomes an empty.
We need to collect maximum number of cherries, using these robots.
Robots can move according to following rules:
\begin{itemize}
\vspace{-5px}
\item[(r1)]
From cell $(i,j)$, robots can move to cell $(i+1, j-1)$, $(i+1, j)$ or $(i+1, j+1)$;
\vspace{-5px}
\item[(r2)]
When both robots stay on the same cell, only one of them takes the cherries;
\vspace{-5px}
\item[(r3)]
Both robots cannot move outside of the grid at any moment;
\vspace{-5px}
\item[(r4)]
Both robots should reach the bottom row in the grid.
\end{itemize}

The fastest solutions, found by us on the network, have $\O(\H{\cdot}\W{\cdot}\min\{\H,\W\})$ complexity.
Same complexity can be reached using next naive DPM with 3D structure $dp$:
for each $i=0,...,\H{-}2$ and $0\le j_1<j_2\le\W-1$
$$
dp[i][j_1][j_2]=\max_{j_1-1\le k_1\le j_1+1, j_2-1\le k_2\le j_2+1, 0\le k_1<k_2<\W}
\{dp[i+1][k_1][k_2]+\g_{i,j_1}+\g_{i,j_2}\}
$$
where $dp[\H-1][j_1][j_2]=\g_{\H-1,j_1}+\g_{\H-1,j_2}$.

Thus, if $2\H>\W$, then we can to find this $dp$ table and return $dp[0][0][\W{-}1]$.
If $2\H\le\W$, then any paths that started from $(0,0)$ and $(0,\W-1)$ don't intersect with each other, then this case can be reduced to the original problem with one path.

Here we answer the question -- is there a solution of \CP\ with $\O(\H{\cdot}\W)$ complexity?
%Also, we reduce DM problem to \CP\ in linear time (without proof of correctness).
Also, we show LS for some extension of \CP\ (without strong proof of correctness).

\subsection{Near linear solution using Suurballe's algorithm}
Here we show the simple reduction of \CP\ to the well known method
for finding 2 node-DP in a edge-weighted digraph (without proof of correctness).

SA is an algorithm for finding 2 node-DP in a nonnegatively-weighted (edge-weighted) digraph, such that both paths connect the same pair of nodes and have minimum total weight.

Let $m$ is maximum value of $\g$.
Denote by $\g'$ the edge-weighted DAG with $\W{\cdot}\H{+}2$ nodes and $3(\W{-}2)(\H{-}1)+4(\H{-}1){+}\W{+}2$ links (directed edges) such that:

1) Each cell of $\g$ contains one node of $\g'$.
And 2 more nodes $s$ and $t$.

2) Weight of link from node in cell $(i,j)$ to node in cell $(i+1,j')$ is $m-\g_{i,j}$,
for each $0\le i<\H-1$, $0\le j<\W$ and $\max\{0, j-1\}\le j'\le\min\{j+1,\W-1\}$.
Weights of 2 links from node $s$ to nodes in cells $(0,0)$ and $(0,\W-1)$ are $0$.
And weight of link from node in cell $(H-1,j)$ to $t$ is $m-\g_{H-1,j}$ for each $0\le j<\W$.

\vspace{4px}
Now we can find 2 node-DP from $s$ to $t$ in $\g'$ using SA.
The total weight of found 2 paths is minimum sum $M'$.
%which can be found by $\O(\H)$.
Then required answer is $m{\cdot}\H-M'$.

\subsubsection{Complexity analysis}

Let denote the set of edges and nodes of graph $\g'$ as $E(\g')$ and $V(\g')$ respectively.
The case when $\W>2\cdot\H$ is trivial (because of this case can be reduced to problem
with one robot in linear time), therefore we can assume that $\W\le 2\cdot\H$.

Complexity of SA equal to complexity of Dijkstra's algorithm (DA) ~\cite{bib:Dijkstra}.
As published in ~\cite{bib:FIBOHEAPOPT} by Fredman and Tarjan the DA can be improved using
Fibonacci heap and performed in $\O(|E(\g')|+|V(\g')|\log(|V(\g')|))$.
Then we get complexity $\O(\H{\cdot}\W{\cdot}\log(\H{\cdot}\W))=\O(\H{\cdot}\W{\cdot}\log(\H))$.

%This reduction doesn't give us such good complexity when we will extend
%\CP\ by other output degree.

There are other optimisations of DA for our purposes.
One of them is algorithm of shortest path (SP) on DAGs.
Using topological sorting we can find SP on DAG in linear time as in ~\cite{bib:SHRT_PTH_INDAG}.
But SA uses search of SP twice.
And before second search of SP, the graph is not a DAG in common case.

Other optimisation for bounded integers weights by some value $C$.
But all such optimisations is not linear.
Most fast of them, in our case, published in ~\cite{bib:SHRT_PTH} by
Ahuja, Mehlhorn, Orlin and Tarjan (AMOT).
This algorithm works in
$\O(|E(\g')|\cdot\log\log(C))$ time.
Thus, if $C$ has polynomial dependence on $H$, then SA with AMOT optimisation has complexity $\O(\H{\cdot}\W{\cdot}\log\log(H))$.

Here we offer linear solution when almost all absolute values of the
grid $\g$ are close to  $C$.

\section{Defaults}

We can assume that absolute values in cells of grids are bounded by value $C=f(\H)$,
for some positive real function $f$ (i.e. in common case some values of $\g$ can be negative).
Exception for values equals to $-\infty$ -- this value is used for bounding of \paths.

Also we assume that $\Theta(\H{\cdot}\W)$ of cells have values $\Theta(f(\H))$.
I.e. the length of input data is $\O(\H{\cdot}\W{\cdot}\log(f(\H)))$.
And assume that $\H,\W\ge2$.

% F = classic DP
\begin{Def}
The $\F_{i,j}(g')$ is table, defined by grid $g'$ of size $\H\times\W$, such that
$$
\F_{i,j}(g') = \begin{cases}
0 &i=\H,\\
g'_{i,j} + \max\{\F_{i+1,\max\{j-1,0\}}(g'), \F_{i+1,j}(g'), \F_{i+1,\min\{j+1,\W-1\}}(g')\} &i=0,...,\H-1
\end{cases}
$$
for each $0\le j\le\W-1$.
By default $\F_{i,j}$ means $\F_{i,j}(\g)$
\end{Def}

% track
\begin{Def}
By {\textbf \path} we call an ordered finite sequence (vector) of cells in grid (by default in $\g$) using rules (r1) and (r3).
I.e.
after not the last cell $(i,j)$ the next cell either $(i+1, \max\{j-1,0\})$ or $(i+1, j)$ or $(i+1, \min\{j+1,\W-1\})$.

Location of \path\ in grid can be obtained by addressing to row number.
For example, at $i$-th row the \path\ $t$ located at $t(i)$-th column.
\end{Def}

% PS
\begin{Def}
Let $t$ is \path\ from row $i_1$ to row $i_2$ ($i_1\le i_2$), then denote sum of $t$ as $\PS(t)$. I.e.
$$
\PS(t) = \sum_{k=i_1}^{i_2}{g_{k,t(k)}}.
$$
\end{Def}

Table $\F$ is known dynamic programming method of search for maximum (or minimum, if we change the $\max$ to $\min$ in the definition of $\F$) sum of falling path.
Also, using $\F$ we can choose one of these \paths\ with maximum sum.

% defined by F
\begin{Def}
Call \path\ $t$ as {\textbf{\path\ defined by $\F_{i',j'}$}}
if $t(i')=j'$ and for each $i{=}i'+1,...,\H{-}1$
$$
	t(i) \in \argmax_{j=\max\{t(i{-}1)-1,0\},...,\min\{t(i{-}1)+1,\W{-}1\}}\{\F_{i,j}\}.\\
$$
\end{Def}

Since the $\F$ is well known DPM for solution of MFPS, then next simple notes we will not prove

\begin{Note} \label{path_defined_by_F_with_sum}
$t$ defined by $\F_{i,j}(\g)$ iif $\PS(t)=\F_{i,j}(\g)$.
\end{Note}

\begin{Note} \label{path_defined_by_F_is_max}
If $t$ starts from cell $(i,j)$ then $\PS(t)\le\F_{i,j}(\g)$.
\end{Note}

% ltrack and rtrack
\begin{Def}
$\lpath$ is leftmost \path\ defined by $\F_{0,0}$.
I.e. $\lpath(0){=}0$ and for each $i{=}1,...,\H{-}1$
$$
	\lpath(i) = \min\argmax_{j=\max\{\lpath(i{-}1)-1,0\},...,\min\{\lpath(i{-}1)+1,\W{-}1\}}\{\F_{i,j}\}.\\
$$
And $\rpath$ is rightmost \path\ defined by $\F_{0,W{-}1}$.
I.e. $\rpath(0) = \W{-}1$ and for each $i=1,...,\H{-}1$
$$
	\rpath(i) = \max\argmax_{j=\max\{\rpath(i{-}1)-1,0\},...,\min\{\rpath(i{-}1)+1,\W{-}1\}}\{\F_{i,j}\}.\\
$$
\end{Def}

By Note \ref{path_defined_by_F_with_sum} we get $\PS(\lpath)=\F_{0,0}(\g)$ and $\PS(\rpath)=\F_{0,\W-1}(\g)$.
Then, if $\lpath$ don't intersect with $\rpath$, then required answer is $\F_{0,0}+\F_{0,\W-1}$.
This case can be checked in $O(\H\times\W)$ of linear operations with numbers of length $\log(\H)$.
Further we suppose that $\lpath$ intersects with $\rpath$.

Due to simmetry of rules by left and right for input data and moving, all properties we will formulate for one side only.
For other side all these properties can be formulated and proved similarly.

By default, if name of pair of \paths\ starts from letters $"l"$ and $"r"$,
then it means that \path\ with first letter $"l"$ located on the left side of \path\ with first letter $"r"$.

When we talk "for each $i$" for rows, we mean "for each $i=0,...,\H-1$".
When we talk "for each $j$" for columns, we mean "for each $j=0,...,\W-1$".

\section{Definitions and properties}

% subtrack
\begin{Def}
Let $0\le i_1<i_2\le\H-1$ and \track\ $t$ with begining not after $i_1$-th row and with ending not before $i_2$-th row.
By {\textbf\subtrack} between rows $i_1$ and $i_2$ of $t$ we call
\track\ $((i_1,t(i_1)),(i_1+1,t(i_1+1)),...,(i_2,t(i_2)))$ and denote it
as $t[i_1,...,i_2]$.

By default $i_1=0, i_2=\H-1$.
\end{Def}
% tail and prefix
\begin{Def}
Let $t$ is \path\ from row $i_1$ to row $i_2$.
By {\textbf{tail}} of \path\ $t$ from $(i,t(i))$ (or from $i$-th row) we call \subpath\ $t[i,...,i_2]$ and denote as $t[i,...]$.

By {\textbf{prefix}} (or head) of \path\ $t$ with end on $(i,t(i))$ we call \subpath\ $t[i_1,...,i]$ and denote $t[...,i]$.
\end{Def}
% concatenation
\begin{Def}
Let $t_1$ and $t_2$ are \paths. Suppose that $t_1$ ends after $(i-1)$-th row
and $t_2$ starts before $(i+2)$.
By {\textbf{\concat\ $t$ of $t_1[...,i]$ and $t_2[i+1,...]$}} we call the sequence of cells ordered by rows where
$t[...,i]=t_1[...,i]$ and $t[i+1,...]=t_2[i+1,...]$.
\end{Def}

\begin{Note} \label{concat_paths}
Let $t_1$ and $t_2$ are \paths\ and $t_1(i)=t_2(i)$
then \concat\ $t$ of $t_1[...,i]$ and $t_2[i+1,...]$ is \path.
I.e. $t$ satisfy the rules (r1) and (r3).
\end{Note}

% cell intersection
\begin{Def}
The {\textbf{\path\ $t$ intersect cell $(k,m)$}} when $t(k)=m$.
\\
The {\textbf{\path\ $t_1$ intersects \path\ $t_2$ at $i$-th row}} when either
$(t_1(i-1)\le t_2(i-1)$ and $t_1(i)\ge t_2(i))$ or
$(t_1(i-1)\ge t_2(i-1)$ and $t_1(i)\le t_2(i))$.
\end{Def}
Note that "paths without intersection" (PWOI) is more stronger than "node-disjoint paths" (or "cell-disjoint paths" in our case).

% swapable tails
\begin{Prop} \label{swapable_tails}
Let \path\ $p_1$ intersects the path $p_2$ at row $i{+}1$ where $p_1(i)\le p_2(i)$ and $p_1(i{+}1)\ge p_2(i{+}1)$,
then tails of $p_1$ and $p_2$ from row $i+1$ are swapable.
It mean that \concat\ of $p_1[0,...,i]$ and $p_2[i+1,...]$ is \path,
and \concat\ of $p_2[...,i]$ and $p_1[i+1,...]$ is \path\ too.
\begin{proof}
%WLOG let prove for concatenation of $p_1[0,...,i]$ and $p_2[i+1,...]$ only.
There are 2 case of intersections:
\begin{itemize}
\item
When $p_1(i)=p_2(i)$.

Then using rule (r1) we get $p_1(i)-1=p_2(i)-1\le p_2(i+1)\le p_2(i)+1=p_1(i)+1$.

I.e. $p_1(i)-1\le p_2(i+1)\le p_1(i)+1$.
Thus $p_1[...,i]$ can be contiued by $p_2[i+1,...]$ without breaking of rule (r1).
A similar proof for \concat\ of $p_2[...,i]$ and $p_1[i+1,...]$.
\item
When $p_1(i)<p_2(i)$.

Then using rule (r1) we get $p_1(i)-1<p_2(i)-1\le p_2(i+1)\le p_1(i+1)\le p_1(i)+1$.
And again, $p_1[...,i]$ can be contiued by $p_2[i+1,...]$ without breaking of rule (r1).

Also using (r1) we get $p_2(i)-1\le p_2(i+1)\le p_1(i+1)\le p_1(i)+1<p_2(i)+1$.
Thus $p_2[...,i]$ can be continued by $p_1[i+1,...]$ without breaking of rule (r1).
\end{itemize}
Since $p_1$ and $p_2$ satisfy the rule (r3),
then any subpaths of them are satisfy the rule (r3).

Thus all these \concats\ satisfy the rules (r1) and (r3).
I.e. \concat\ of $p_1[0,...,i]$ and $p_2[i+1,...]$ is \path,
and \concat\ of $p_2[...,i]$ and $p_1[i+1,...]$ is \path\ too.
\end{proof}
\end{Prop}

\begin{Note} \label{tail_of_max}
If \path\ $t$ defined by $\F_{i,j}$, then for any row $i'\ge i$ we get $\PS(t[i',...])=\F_{i',t(i')}$.
\end{Note}

% swapable maximum subtracks
\begin{Prop} \label{swapable_max}
Consider \path\ $t_1$ started from cell $(i_1,j_1)$ and has maximum sum (i.e. $t_1$ is \path\ defined by $\F_{i_1,j_1}$).
Suppose that $t_1$ intersect $(k_1,m_1)$-th and $(k_2,m_2)$-th cells, where $k_2>k_1\ge i_1$. Then:
\vspace{-4px}
\begin{enumerate}
\item
$\PS(t_1[k_1,...,k_2-1])=\F_{k_1,m_1}-\F_{k_2,m_2}$;
\vspace{-4px}
\item
Let \path\ $t$ intersect cells $(k_1,m_1)$ and $(k_2,m_2)$ then $\PS(t[k_1,...,k_2])\le \PS(t_1[k_1,...,k_2])$;
\vspace{-4px}
\item
Let \path\ $t$ intersect cell $(k_1,m_1)$ and $t$ intersect $t_1$ at row $k_2$ then $\PS(t[k_1,...,k_2-1])\le \PS(t_1[k_1,...,k_2-1])$;
\vspace{-4px}
\item
Let \path\ $t$ intersect cells $(k_1,m_1)$ and $(k_2,m_2)$,
and $\PS(t[k_1,...,k_2-1])=\F_{k_1,m_1}-\F_{k_2,m_2}$.
Then for any $k_1\le k_1'\le k_2'\le k_2$ we get $\PS(t[k_1',...,k_2'-1])=\F_{k_1',t(k_1')}-\F_{k_2',t(k_2')}$;
\item
Let \path\ $t$ intersect cells $(k_1,m_1)$ and $(k,m)$ for some $k>k_1$ and $0\le m\le\W-1$,
then $\PS(t[k_1,...,k-1])\le\F_{k_1,m_1}-\F_{k,m}$.
\end{enumerate}
\begin{proof}
\begin{enumerate}
\item
Since $t_1$ defined by $\F$, then for any row $i\ge i_1$
%the tail $t_1[i,...]$ has maximum sum among all \paths\ started from cell $(i,t_1(i))$
by Note \ref{tail_of_max} we get $\PS(t_1[i,...])=\F_{i,t_1(i)}$.
Thus $\PS(t_1[k_1,...,k_2-1])=\PS(t_1[k_1,...)-\PS(t_1[k_2,...])=
\F_{k_1,m_1}-\F_{k_2,m_2}$.
\item
Suppose that $\PS(t[k_1,...,k_2])>\PS(t_1[k_1,...,k_2])$.

Let $t'$ is \concat\ with begining on cell $(k_1,m_1)$ such that
$t'[k_1,...,k_2]=t[k_1,...,k_2]$ and $t'[k_2+1,...]=t_1[k_2+1,...]$.
By Note \ref{concat_paths} the $t'$ is \path.

Then $\F_{k_1,m_1}\ge\PS(t')$ and the other side:\\
$\begin{array}{lll}
\PS(t')&=\PS(t[k_1,...,k_2])+\PS(t_1[k_2+1,...])>\\
&>\PS(t_1[k_1,...,k_2])+\PS(t_1[k_2+1,...])=\PS(t_1[k_1,...])=\F_{k_1,m_1}
\end{array}$

This contradiction proves statement 2.
\item
Let $t'$ is \concat\ of $t[k_1,...,k_2-1]$ and $t_1[k_2,...]$.
By Property \ref{swapable_tails} $t'$ is \path.
Also $t'$ intersects with cells $(k_1,m_1)$ and $(k_2,m_2)$.
Then using Property \ref{swapable_max}.2 we get
$\PS(t_1[{k_1,...,k_2-1}])=\PS(t_1[k_1,...,k_2])-\g_{k_2,m_2}\ge\PS(t'[k_1,...,k_2])-\g_{k_2,m_2}=\PS(t[{k_1,...,k_2-1}])$.
\item
Let $t_2$ is \path\ defined by $\F_{k_2,m_2}$.
And $t'$ is \concat\ of $t[k_1,...,k_2-1]$ and $t_2[k_2,...]$.
Then by Note \ref{concat_paths} $t'$ is \path,
with sum $\PS(t')=\PS(t[k_1,...,k_2-1])+\PS(t_2[k_2,...])=\F_{k_1,m_1}-\F_{k_2,m_2}+\F_{k_2,m_2}=\F_{k_1,m_1}$.
I.e. $t'$ defined by $\F_{k_1,m_1}$.

Then using Property \ref{swapable_max}.1 we get $\PS(t'[k_1',...,k_2'-1])=\F_{k_1',t'(k_1')}-\F_{k_2',t'(k_2')}$.
Since $t(k_2)=t'(k_2)$ then $t[k_1',...,k_2']=t'[k_1',...,k_2']$ then
$\PS(t[k_1',...,k_2'-1])=\F_{k_1',t(k_1')}-\F_{k_2',t(k_2')}$.
\item
Let $b_1=\max\{0,t(k-1)-1\}$ and $b_2=\min\{t(k-1)+1,\W-1\}$.
Then $m\in\{b_1,...,b_2\}$.

Let prove by induction on difference $k-k_1$

{\bf Base case:}\\
If $k-k_1=1$ then $\PS(t[k_1,...,k-1])=\PS(t[k_1])=\g_{k_1,m_1}=\g_{k-1,t(k-1)}\le\\
\le\g_{k-1,t(k-1)}+\max_{j=b_1,...,b_2}\{\F_{k,j}\}-\F_{k,m}
=\F_{k_1,m_1}-\F_{k,m}$.

{\bf Induction step:}\\
Let $k-k_1>1$,
and $\PS(t[k_1,...,k-2])\le\F_{k_1,m_1}-\F_{k-1,t(k-1)}$.

Then $\PS(t[k_1,...,k-1])=\PS(t[k_1,...,k-2])+\g_{k-1,t(k-1)}\le\\
\le\PS(t[k_1,...,k-2])+\g_{k-1,t(k-1)}+\max_{j=b_1,...,b_2}\{\F_{k,j}\}-\F_{k,m}=\\
=\PS(t[k_1,...,k-2])+\F_{k-1,t(k-1)}-\F_{k,m}\le(\F_{k_1,m_1}-\F_{k-1,t(k-1)})+\F_{k-1,t(k-1)}-\F_{k,m}=\\
=\F_{k_1,m_1}-\F_{k,m}$.
\end{enumerate}
\end{proof}
\end{Prop}

% ltrack <= rtrack
\begin{Note} \label{lr_relation}
$\lpath(i)\le\rpath(i)$ for each $i=0,...,\H-1$.
\end{Note}
% PS(ltrack)=F(0,0)
\begin{Note} \label{ltrack_F00}
$\PS(\lpath)=\F_{0,0}$ and $\PS(\rpath)=\F_{0,\W-1}$.
\end{Note}

% lg and rg tables
\begin{Def}
$\lg$ is grid defined for each $i=0,...,\H-1$ as:
$$
\lg_{i,j} = \begin{cases}
-\infty  &j=\lpath(i)+1,...,\W-1,\\
\g_{i,j} &j=0,...,\lpath(i).
\end{cases}
$$
And $\rg$ is grid defined for each $i=0,...,\H-1$ as:\\
$$
\rg_{i,j} = \begin{cases}
\g_{i,j} &j=\rpath(i),...,\W-1,\\
-\infty  &j=0,...,\rpath(i)-1.
\end{cases}
$$
\end{Def}

\begin{Prop} \label{Flg_is_Fg_for_jleltrack}
For each $i=0,...,\H-1$ and $j\le\lpath(i)$ we get $\F_{i,j}(\g)=\F_{i,j}(\lg)$,
and for $j\ge\rpath(i)$ we get $\F_{i,j}(\g)=\F_{i,j}(\rg)$.
\begin{proof}
Due to $\g_{i,j}\ge\lg_{i,j}$ for each $i$ and $j$,
we get $\F_{i,j}(\g)\ge\F_{i,j}(\lg)$ for each $i$ and $j$.

Let $t$ is $\path$ $defined$ $by$ $\F_{i_1,j_1}(\g)$ for some $i_1$ and $j_1\le\lpath(i_1)$, then $\PS(t)=\F_{i_1,j_1}(\g)$.

Consider 2 cases:
\begin{itemize}
\item
If $t(i)\le\lpath(i)$ for each $i$,
then $\F_{i_1,j_1}(\lg)\ge\PS(t)=\F_{i_1,j_1}(\g)$.% which contradicts with our assumption.
\item
Let $i_2$ is lowest row such that $t(i_2)>\lpath(i_2)$ (i.e. $i_2>i_1$).
Then due to Property \ref{swapable_tails} a \concat\ $t'$ of $\lpath[...,i_2-1]$ and $t[i_2,...]$ is \path.

Since $t$ defined by $\F(\g)$ then by Note \ref{tail_of_max} we get
$\PS(t[i_2,...])=\F_{i_2,t(i_2)}(\g)$.
Since $\lpath$ defined by $\F(\g)$ then by Property \ref{swapable_max}.1 we get
$\PS(\lpath[...,i_2-1])=\F_{0,0}(\g)-\F_{i_2,\lpath(i_2)}(\g)$.

Then
$\F_{0,0}(\g)\ge\PS(t')=\PS(\lpath[...,i_2-1])+\PS(t[i_2,...])=
\F_{0,0}(\g)-\F_{i_2,\lpath(i_2)}(\g)+\F_{i_2,t(i_2)}(\g)$.
Thus $\F_{i_2,\lpath(i_2)}(\g)\ge\F_{i_2,t(i_2)}(\g)$.

Consider \concat\ $t''$ of
$t[i_1,...,i_2-1]$ and $\lpath[i_2,...]$.
Then due to Property \ref{swapable_tails} the $t''$ is \path.

Since $\lpath$ defined by $\F(\g)$, due to Note \ref{tail_of_max} we get
$\PS(\lpath[i_2,...])=\F_{i_2,\lpath(i_2)}(\g)$.
By {Property \ref{swapable_max}.1} we get
$\PS(t[i_1,...,i_2-1])=\F_{i_1,j_1}(\g)-\F_{i_2,t(i_2)}(\g)$.
Then\\
$\begin{array}{ll}
\PS(t'')&=\PS(t[i_1,...,i_2{-}1])+\PS(\lpath[i_2,...])
=\F_{i_1,j_1}(\g)-\F_{i_2,t(i_2)}(\g)+\F_{i_2,\lpath(i_2)}(\g)\ge\\
&\ge\F_{i_1,j_1}(\g).
\end{array}$
\\
By our choice of $t'$ we get $t''(i)\le\lpath(i)$ for each $i$.
Then $\F_{i_1,j_1}(\lg)\ge\PS(t'')\ge\F_{i_1,j_1}(\g)$.
\end{itemize}
Similarly we can proof that $\F_{i,j}(\g)=\F_{i,j}(\rg)$.
\end{proof}
\end{Prop}

% maximum of each interval of ltrack and rtrack
\begin{Prop} \label{l_max}
Let $0{\le} i_1{<}i_2{\le}\H{-}1$, and
consider \path\ $t$ from cell $(i_1,\lpath(i_1))$ to cell $(i_2,\lpath(i_2))$,
and \path\ $t'$ from cell $(i_1,\rpath(i_1))$ to cell $(i_2,\rpath(i_2))$.
Then:
\begin{enumerate}
\item
Due to Property \ref{swapable_max}.2 and Note \ref{ltrack_F00} we get $\PS(t)\le \PS(\lpath[i_1,...,i_2])$.\\
Similarly we get $\PS(t')\le \PS(\rpath[i_1,...,i_2])$.
\item
Due to Property \ref{l_max}.1, leftmost of $\lpath$ and rigthmost of $\rpath$ we get implication:\\
if $\PS(t)=\PS(\lpath[i_1,...,i_2])$
then $t(i)\ge\lpath(i)$ for each $i=i_1,...,i_2$;\\
if $\PS(t')=\PS(\rpath[i_1,...,i_2])$
then $t'(i)\le\rpath(i)$ for each $i=i_1,...,i_2$.
\item
If $t$ is $\LP$ \path\ and $\PS(t)= \PS(\lpath[i_1,...,i_2])$, then by Property \ref{l_max}.2 we get $t=\lpath[i_1,...,i_2]$.\\
Similarly, if $t'$ is $\RP$ \path\ and $\PS(t')= \PS(\rpath[i_1,...,i_2])$, then $t'=\rpath[i_1,...,i_2]$.
\item
If $p$ is $\LP_{i_1,\lpath(i_1)}$ \path\ and $\PS(p)= \PS(\lpath[i_1,...])$, then due to leftmost and maximum sum of $\lpath$ we get $p=\lpath[i_1,...]$.\\
Similarly, if $p'$ is $\RP_{i_1,\rpath(i_1)}$ \path\ and $\PS(p')= \PS(\rpath[i_1,...])$, then $p'=\rpath[i_1,...]$.
\end{enumerate}
\end{Prop}
% LP and RP
\begin{Def}
Let \path\ $t$ with begining at cell ${(i,j)}$ and ends at $i'$-th row.

If ${t(k) \le \lpath(k)}$ for each ${k=i,...,i'}$ then call $t$ as $\LP_{i,j}$ \path.

If ${t(k) \ge \rpath(k)}$ for each ${k=i,...,i'}$ then call $t$ as $\RP_{i,j}$ \path.
\end{Def}
% ltrack <= rtrack
\begin{Note} \label{l_RP}
If $t$ is $\LP$ \path, and $t(i)=\rpath(i)$, then $\lpath(i)=\rpath(i)$.
If $t$ is $\RP$ \path, and $t(i)=\lpath(i)$, then $\lpath(i)=\rpath(i)$.
\end{Note}
% inherition of unintersections
\begin{Note} \label{concat_without_intersection}
Let \paths\ $t_1$,..., $t_n$ don't intersect the \path\ $t_0$,
and all $t_1$,..., $t_n$ are placed on the same side of $t_0$.
And $t$ is \concat\ of $t_1,...,t_n$ \subpaths, such that $t$ is \path.
Then $t$ is \path\ without intersection with any \subpath\ of $t_0$.
\end{Note}
% inherition of RP
\begin{Note} \label{LRP_inherits}
Let $t_1$,..., $t_n$ are $\RP_{i_1,t_1(i_1)}$, ..., $\RP_{i_n,t_n(i_n)}$
\tracks\ respectively, and $t$ is \concat\ of $t_1$,..., $t_n$ \subpaths,
such that $t$ is \path.
Then $t$ is $\RP_{i,j}$ \path\ for some $i$ and $j\ge\rpath(i)$.

Let $t_1$,..., $t_n$ are $\LP_{i_1,t_1(i_1)}$, ..., $\LP_{i_n,t_n(i_n)}$
\tracks\ respectively, and $t$ is \concat\ of $t_1$,..., $t_n$ \subpaths,
such that $t$ is \path.
Then $t$ is $\LP_{i,j}$ \path\ for some $i$ and $j\le\lpath(i)$.
\end{Note}
% pairs with maximum summ (lrtms{-}pair(i,j1,j2))
\begin{Def}
Let $t_1$ and $t_2$ are $\LP_{i,j_1}$ and $\RP_{i,j_2}$ PWOI, such that $\PS(t_1)+\PS(t_2)$ is maximum among all $\LP_{i,j_1}$ and $\RP_{i,j_2}$ pairs of PWOI and ending at bottom (BE), then we call this pair as {\bf pair with maximum sum}, and denote as {\bf $\lrtms(i,j_1,j_2)$ pair} ((l)eft and (r)ight (d)isjoint (t)racks with (m)aximum (s)um).
\end{Def}

\begin{Def}
$\rMaxes$ is table, where $\rMaxes(i,j)=\PS(l)+\PS(r)$ for any $\lrtms(i,j,\rpath(i))$ pair $l$ and $r$.
I.e. $\rMaxes(i,j)$ is maximum sum among all pairs of $\LP_{i,j}$ and $\RP_{i,\rpath(i)}$ PWOI and BE.

$\lMaxes$ is table, where $\lMaxes(i,j)=\PS(l)+\PS(r)$ for any
$\lrtms(i,\lpath(i),j)$ pair $l$ and $r$.
\end{Def}

\begin{Note} \label{M_defined}
For each row $i$ the $\rMaxes$ defined in columns $j\le\min\{\lpath(i), \rpath(i)-1\}$ only.
For each row $i$ the $\lMaxes$ defined in columns $j\ge\max\{\lpath(i)+1, \rpath(i)\}$ only.
\end{Note}

\subsection{Linear search of $\lMaxes$ and $\rMaxes$}

% equality of lPmax to ltrack between intersections
\begin{Prop} \label{between_intersections}
Let $lt$ and $rt$ are $\lrtms(i,j_1,j_2)$ pair, for some $j_1\le\lpath(i)$ and $j_2\ge\rpath(i)$.
\begin{enumerate}
\item
If $lt$ intersect $\lpath$ at 2 rows $i_2>i_1>i$, and $rt$ don't intersect $\lpath$ between these rows, then $lt[i_1,...,i_2]=\lpath[i_1,...,i_2]$.
\item
If $lt$ intersect $\lpath$ at row $i'$, and $rt$ don't intersect $\lpath$ after this row, then $lt[i',...]=\lpath[i',...]$.
\end{enumerate}
\begin {proof}
\begin{enumerate}
\item
Suppose that $lt[i_1,...,i_2]\ne\lpath[i_1,...,i_2]$.

If suppose that $\PS(lt[i_1,...,i_2])=\PS(\lpath[i_1,...,i_2])$ then by Property \ref{l_max}.3 we get
$lt[i_1,...,i_2]=\lpath[i_1,...,i_2]$ that contradicts to our assumption.
Thus, using Property \ref{l_max}.1, we get inequality $\PS(lt[i_1,...,i_2])<\PS(\lpath[i_1,...,i_2])$.

Since $lt$ is $\LP$ \path\ then because of the intersection with $\lpath$ on $i_1$ and $i_2$
we get $lt(i_1)=\lpath(i_1)$ and $lt(i_2)=\lpath(i_2)$.
Then consider \concat\ $lt'$:\\
$\begin{array}{lll}
\ \ \ \ \ \ \ \ &lt'[i,...,i_1-1]&=lt[i,...,i_1-1],\\
&lt'[i_1,...,i_2]&=\lpath[i_1,...,i_2],\\
&lt'[i_2+1,...]&=lt[i_2+1,...].
\end{array}$

By Note \ref{concat_paths} the $lt'[i_1,...]$ is \path.
Then by Note \ref{concat_paths} the $lt'$ is \path.
By Note \ref{LRP_inherits} the $lt'$ is $\LP_{0,0}$ \path.
By Note \ref{concat_without_intersection} $lt'$ don't intersects with $rt$.

Consider relation between $\PS(lt)$ and $\PS(lt')$:\\
$\begin{array}{lll}
\PS(lt)&=\PS(lt[i,...,i_1{-}1])+\PS(lt[i_1,...,i_2])&+\PS(lt[i_2{+}1,...])<\\
&<\PS(lt[i,...,i_1{-}1])+\PS(\lpath[i_1,...,i_2])&+\PS(lt[i_2{+}1,...])=\PS(lt').
\end{array}$

Thus we get $lt'$ and $rt$ are $\LP_{i,j_1}$ and $\RP_{i,j_2}$ \tracks\ without intersection
with sum $\PS(lt')+\PS(rt)>\PS(lt)+\PS(rt)$. That contradict to maximum sum of $\lrtms(i,j_1,j_2)$ pair $lt$ and $rt$.
\item
Suppose that $lt[i',...]\ne\lpath[i',...]$.
Since $lt$ is $\LP$ \path\ then because of the intersection with $\lpath$ on $i'$
we get $lt(i')=\lpath(i')$ and $lt[i'+1,...]\ne\lpath[i'+1,...]$.

Then consider \concats\ $lt'$ and $lt''$:\\
$\begin{array}{lllll}
&lt'[...,i']&=lt[...,i'], &lt'[i'+1,...]&=\lpath[i'+1,...]\\
&lt''[...,i']&=\lpath[...,i'], &lt''[i'+1,...]&=lt[i'+1,...].
\end{array}$

By Note \ref{concat_paths} the $lt'$ and $lt''$ are \paths.
Then by Note \ref{LRP_inherits} the $lt'$ and $lt''$ are $\LP$ \paths.
By Note \ref{concat_without_intersection} $lt'$ don't intersects with $rt$.

Since $lt''[i'+1,...]=lt[i'+1,...]\ne\lpath[i'+1,...]$ then $lt''\ne\lpath$.
Then due to leftmost of $\lpath$ among all $\LP$ \paths\ with maximum sum we get
$\PS(\lpath)>\PS(lt'')$.
Then\\
$\PS(lt[i'{+}1,...])=\PS(lt'')-\PS(\lpath[...,i'])<\PS(\lpath)-\PS(\lpath[...,i'])=
\PS(\lpath[i'{+}1,...])$.

Then $\PS(lt)=\PS(lt[...,i'])+\PS(lt[i'+1,...])<\PS(lt[...,i'])+\PS(lpath[i'+1,...])=\PS(lt')$.

Thus we get $\LP_{i,j_1}$ and $\RP_{i,j_2}$ \paths\ $lt'$ and $rt$ without intersections
with sum $\PS(lt')+\PS(rt)>\PS(lt)+\PS(rt)$.
That contradict to maximum sum of $\lrtms(i,j_1,j_2)$ pair $lt$ and $rt$.
\end{enumerate}
\end {proof}
\end{Prop}

\begin{Prop} \label{tail_of_lrtms_is_lrtms}
Let $lt$ and $rt$ are $\lrtms(i,j_1,j_2)$ pair.
Then for any $i'\ge i$ the pair $lt[i',...]$ and $rt[i',...]$ are
$\lrtms(i',lt(i'),rt(i'))$ pair.
\begin{proof}
By Note \ref{concat_without_intersection} the $lt[i',...]$ don't intersects with $rt[i',...]$.
By Note \ref{LRP_inherits} the $lt[i',...]$ and $rt[i',...]$ are
$LP_{i',lt(i')}$ and $RP_{i',rt(i')}$ \paths\ respectively.

Let $lmt$ and $rmt$ are $\lrtms(i',lt(i'),rt(i'))$ pair.
Suppose that $\PS(lmt)+\PS(rmt)>\PS(lt[i',...])+\PS(rt[i',...])$.
Consider \concats\ $lp$ and $rp$ such that:\\
$\begin{array}{lllll}
&lp[i,...,i'-1]&=lt[i,...,i'-1], &lp[i',...]&=lmt[i',...],\\
&rp[i,...,i'-1]&=rt[i,...,i'-1], &rp[i',...]&=rmt[i',...].\\
\end{array}$\\
By Note \ref{concat_paths} the $lp$ and $rp$ are \paths.
By Note \ref{LRP_inherits} $lp$ is $LP_{i,j_1}$ \path\ 
and $rp$ is $RP_{i,j_1}$ \path.

Since $lt[i,...,i'-1]$ don't intersects with $rt[i,...,i'-1]$, and
$lmt[i',...]$ don't intersects with $rmt[i',...]$,
then $lp$ don't intersects with $rp$.
Then due to maximum sum of $lt$ and $rt$ we get $\PS(lp)+\PS(rp)\le\PS(lt)+\PS(rt)$.
But the other side\\
$\begin{array}{ll}
\PS(lp)+\PS(rp)&=\PS(lt[i,...,i'-1])+\PS(lmt[i',...])+\PS(rt[i,...,i'-1])+\PS(rmt[i',...])>\\&>\PS(lt[i,...,i'-1])+\PS(lt[i',...])+\PS(rt[i,...,i'-1])+\PS(rt[i',...])=\\
&=\PS(lt)+\PS(rt).
\end{array}$\\
This contradiction proves that
$\PS(lmt)+\PS(rmt)=\PS(lt[i',...])+\PS(rt[i',...])$.

Thus we get $LP_{i',lt(i')}$ and $RP_{i',rt(i')}$ \paths\ 
$lt[i',...]$ and $rt[i',...]$ respectively without intersection with maximum sum.
I.e. $lt[i',...]$ and $rt[i',...]$ are $\lrtms(i',lt(i'),rt(i'))$ pair.
\end{proof}
\end{Prop}

\begin{Prop} \label{variable_max_prefix}
Let $lt$ and $rt$ are $\lrtms(i,lt(i),rt(i)))$ pair,
$lt[i,...,ri]$ don't intersects with $\lpath[i,...,ri]$
and $rt(ri)=\lpath(ri)$ for some $i<ri$.
Let $i<i'<ri$ and $\rpath(i')\le j'\le rt(i')$.
Consider $\RP_{i',j'}$ \path\ $rt'$ where $rt'[ri,...]=rt[ri,...]$
and $\PS(rt'[i',...,ri])=\F_{i',rt'(i')}-\F_{ri,rt'(ri)}+\g_{ri,rt'(ri)}$.
Then $lt[i',...]$ and $rt'$ are $\lrtms(i',lt(i'),j')$ pair.
\begin{proof}
Since $lt$ is $\LP_{i,lt(i)}$ path and don't intersects with $\lpath[i,...,ri]$, then
$lt(k)<\lpath(k)\le\rpath(k)\le rt'(k)$ for each $k=i',...,ri$.
Since $lt$ don't intersects with $rt$,
then by Note \ref{concat_without_intersection} the $lt[i',...]$ don't intersects with $rt'$.

Let denote $lt[i',...]$ and $rt[i',...]$ as $lT$ and $rT$ respectively.
Consider $\lrtms(i',lt(i'),j')$ pair $lP$ and $rP$.
Since $rP$ is $\RP_{i',j'}$ \path\ and $j'\le rt(i')=rT(i')$, then
$rP$ intersects with $rT$ on some row $rI\le ri$.
Let $rI$ is first row of intersection of $rP$ and $rT$.
Then $rT$ don't intersects with $lP$ before $rI$.
Since $lT$ don't intersects with any of $\RP$ \path\ before $ri$, then $lT$ don't intersects with $rP$ before $rI$.

Let $rP_1$ and $rP'$ are \concats:\\
$\begin{array}{lllll}
&rT'[i',...,rI-1]&=rP[i',...,rI-1], &rT'[rI,...]&=rT[rI,...],\\
&rP'[i',...,rI-1]&=rT[i',...,rI-1], &rP'[rI,...]&=rP[rI,...].
\end{array}$\\
If $rP(rI)=rT(rI)$ then by Note \ref{concat_paths} the $rT'$ and $rP'$ are \paths.
If $rP(rI)\ne rT(rI)$ then $rP(rI)>rT(rI)$ then by Property \ref{swapable_tails} the $rT'$ and $rP'$ are \paths.
Then by Note \ref{LRP_inherits} $rT'$ and $rP'$ are $\RP$ \paths.
Using Note \ref{concat_without_intersection} the $lP$ don't intersects with $rP'$, and
$lT$ don't intersects with $rT'$.

Consider relations of differences $d_1=\PS(lP)-\PS(lT)$ and $d_2=\PS(rT[rI,...])-\PS(rP[rI,...])$:
\begin{itemize}
\item
$d_1>d_2$.
We get $\LP_{i', lt(i')}$ and $\RP_{i', rt(i')}$ \paths\ $lP$ and $rP'$
without intersections with sum
\\
$\begin{array}{ll}
\PS(lP)+\PS(rP')&=d_1+\PS(lT)+\PS(rT[i',...,rI-1])+\PS(rP[rI,...])=\\
&=d_1+\PS(lT)+\PS(rT[i',...,rI-1])+\PS(rT[rI,...])-d_2>\\
&>\PS(lT)+\PS(rT).
\end{array}$\\
which conrtadicts to maximum of $\PS(lT)+\PS(rT)$ due to Property \ref{tail_of_lrtms_is_lrtms}.
\item
$d_1\le d_2$.
We get $\LP_{i', lt(i')}$ and $\RP_{i', j'}$ \paths\ $lT$ and $rT'$
without intersections with sum
\\
$\begin{array}{ll}
\PS(lT)+\PS(rT')&=\PS(lP)-d_1+\PS(rP[i',...,rI-1])+\PS(rT[rI,...])=\\
&=\PS(lP)-d_1+\PS(rP[i',...,rI-1])+\PS(rP[rI,...])+d_2\ge\\
&\ge\PS(lP)+\PS(rP).
\end{array}$\\
Inequality $\PS(lT)+\PS(rP_1)>\PS(lP)+\PS(rP)$ conrtadicts the maximum of $\PS(lP)+\PS(rP)$
among all pairs of $\LP_{i', lt(i')}$ and $\RP_{i', j'}$ \paths\ without intersections.
\end{itemize}
Thus we get one valid case $d_1=d_2$ with equation $\PS(lT)+\PS(rT')=\PS(lP)+\PS(rP)$.
I.e. $lT=lt[i',...]$ and $rT'$ are $\lrtms(i',lt(i'),j')$ pair.
Since $rI\le ri$ then $rT'(ri)=rT(ri)=rt(ri)$.

Thus we get $RP_{i', j'}$ \path\ $rT'$ where $rT'[ri,...]=rt[ri,...]$.
Using Properties \ref{Flg_is_Fg_for_jleltrack} and \ref{swapable_max}.5 we get
$
\PS(rT'[i',...,ri])\le\F_{i',j'}-\F_{ri,rT'(ri)}+\g_{ri,rT'(ri)}=\F_{i',rt'(i')}-\F_{ri,rt'(ri)}+\g_{ri,rt'(ri)}=
\PS(rt'[i',...,ri])
$.

Then, using condition $rI\le ri$, we get\\
$\begin{array}{ll}
\PS(lt[i',...])+\PS(rt')&=\PS(lT)+\PS(rt'[i',...,ri])+\PS(rt'[ri+1,...])\ge\\
&\ge\PS(lT)+\PS(rT'[i',...,ri])+\PS(rt[ri+1,...])=\\
&=\PS(lT)+\PS(rT'[i',...,ri])+\PS(rT[ri+1,...])=\PS(lT)+\PS(rT').
\end{array}$

Thus we get that $lt[i',...]$ and $rt'$ are $\LP_{i',lt(i')}$ and $\RP_{i',j'}$ \paths\ respectively without intersections and with maximum sum.
I.e. $lt[i',...]$ and $rt'$ are $\lrtms(i',lt(i'),j')$ pair.
\end{proof}
\end{Prop}

% continuation from nearest intersection of ltrack and rtrack
\begin{Prop} \label{nearest_intersection}
Let $lt$ and $rt$ are $\lrtms(i-1,\lpath(i-1),j)$ pair, where $j>\rpath(i-1)$. And $lt(i)<\lpath(i)$, $rt(i)>\rpath(i)$.
Then:
\begin{enumerate}
\item
Exist $ri>i$ such that $rt(ri)=\lpath(ri)$, and $lt(k)<\lpath(k)$ for each $k=i,...,ri$;
\item
Consider \concat\ $rt'$ of $\rpath[i,...,ri{-}1]$ and $rt[ri,...]$ (i.e. $rt'[...,ri]=\rpath[i,...,ri]$).
Then $lt[i,...]$ and $rt'$ are $\lrtms(i,lt(i),\rpath(i))$ pair;
\item
$\PS(rt[i{-}1,...,ri])=\F_{i{-}1,rt(i{-}1)}-\F_{ri,rt(ri)}+\g_{ri,rt(ri)}$.
And $\PS(rt[i,...,ri{-}1])=\F_{i,rt(i)}-\F_{ri,rt(ri)}$ by Property \ref{swapable_max}.1;
\item
Let $b_1=\max\{0,\lpath(i-1)-1\}, b_2=\min\{\lpath(i-1)+1,\lpath(i)-1\}$ and\\
$b_3=\max\{\rpath(i)+1,j-1\}, b_4=\min\{j+1,\W-1\}$
then
$$\PS(lt[i,...])+\PS(rt[i,...])=\max_{k=b_1,...,b_2}\{\rMaxes(i,k)\}+\max_{k=b_3,...,b_4}\{\F_{i,k}\}-\F_{i,\rpath(i)}.$$
\end{enumerate}
\begin {proof}
\begin{enumerate}
\item
Suppose that $rt$ don't intersect $\lpath$ after $(i{-}1)$-th row.
Then due to Property \ref{between_intersections}.2 we get $lt[i-1,...]=\lpath[i-1,...]$
that contradicts with condition $lt(i)<\lpath(i)$.
I.e. $rt$ intersect $\lpath$ after $(i-1)$-th row.

Let $ri\ge i$ such that $rt(ri)=\lpath(ri)$, and $rt(k)\ne\lpath(k)$ for each $k=i-1,...,ri-1$.

Suppose that $lt$ intersects with $\lpath$ on row $li$ between $i$ and $ri$.
Since $lt(i{-}1)=\lpath(i{-}1)$, then due to Property \ref{between_intersections}
we get $lt[i{-}1,...,li]=\lpath[i{-}1,...,li]$ that contradicts with $lt(i)<\lpath(i)$.

Thus $lt(k)\ne\lpath(k)$ for each $k=i,...,ri$.
Then because of $lt$ is $\LP$ \path\ then $lt(k)<\lpath(k)$ for each $k=i,...,ri$.
Since $\rpath(i)<rt(i)$ then $ri>i$.
\item
Since $rt(ri)=\lpath(ri)$ then using Note \ref{l_RP} we get $rt'[i,...,ri]=\rpath[i,...,ri]$.
Due to Note \ref{concat_paths} the $rt'$ is \path.
By Note \ref{LRP_inherits} the $rt'$ is $\RP_{i,j}$ \path.

Since $\rpath$ defined by $\F$ then
$\PS(rt'[i,...,ri])=\PS(\rpath[i,...,ri])=\F_{i,rt'(i)}-\F_{ri,rt'(ri)}+\g_{ri,rt'(ri)}$.
Then due to Property \ref{variable_max_prefix} the $lt[i,...]$ and $rt'$ are
$\lrtms(i,lt(i),rt'(i))$ pair.
Since $rt'(i)=\rpath(i)$ we get proof of statement 2.
\item
Since $\lpath(i-1)=lt(i-1)$ and $lt$ is $LP$ path then by Property \ref{l_max}.1
$\PS(\lpath[i-1,...])\ge\PS(lt)$.
Since $\lpath(i)<lt(i)$ then $\lpath[i-1,...]\ne lt[i-1,...]$.
Then since $\lpath(i-1)=lt(i-1)$ and $lt$ is $LP$ by Property \ref{l_max}.4 we get
$\PS(\lpath[i-1,...])>\PS(lt[i-1,...])$.

Consider $\RP_{i-1,j}$ \path\ $rt''$ defined by $\F_{i-1,j}(\rg)$.
Then $\PS(rt)\le\F_{i-1,j}(\rg)=\PS(rt'')$.

In case when $rt'(k)<rt''(k)$ for each $k\ge i-1$ we get $\lpath[i-1,...]$ and
$rt''$ are $\LP_{i-1,\lpath(i-1)}$ and $\RP_{i-1,j}$ \paths\ without intersections and with sum
$\PS(\lpath[i-1,...])+\PS(rt'')>\PS(lt)+\PS(rt)$
that contradict to maximum sum of $lt$ and $rt$.
I.e. this case impossible.

Then $rt'(i')\ge rt''(i')$ for some $i'>i-1$.
WLOG we can assume that $rt'(k)<rt''(k)$ for each $k=i-1,...,i'-1$.

Let $ri'>ri$ such that $\rpath(ri')<rt(ri')$ and $\rpath(k)=rt(k)$ for each $k=ri,...,ri'-1$.
I.e. using Property \ref{nearest_intersection}.1 we get $lt(k)<\rpath(k)$ for each $k=i,...,ri'-1$.
And since $rt''(i{-}1)=j>\rpath(i{-}1)\ge\lpath(i{-}1)$ then $lt$
don't intersects with $rt''[i{-}1,...,ri'{-}1]$.
If $rt[ri,...]=\rpath[ri,...]$ then we can assume that $ri'=\H$ and $\F_{\H,k}=0$ for each $k=0,...,\W-1$.

If $i'<ri'$ then $rt''(i')=rt'(i')=\rpath(i')$ then due to Property \ref{l_max}.4
we get $rt''[i',...]=\rpath[i',...]$.
Then $rt''(ri'-1)=\rpath(ri'-1)=rt(ri'-1)$.

Then due to Properties \ref{swapable_max}.3 %and maximum sum of $lt$ and $rt$
we get $\PS(rt''[i-1,...,ri'-2])\ge\PS(rt[i{-}1,...,ri'{-}2])$.

Suppose that $\PS(rt''[i-1,...,ri'-2])>\PS(rt[i-1,...,ri'-2])$.
Then consider \concat\ $rp$ of $rt''[i-1,...,ri'-2]$ and $rt[ri'-1,...]$.
Since $rt''(ri'-1)=rt(ri'-1)$ then by Property \ref{swapable_tails} the $rp$ is \path.
By Note \ref{LRP_inherits} the $rp$ is $\RP$ \path.
By Note \ref{concat_without_intersection} the $lt$ don't intersects with $rp$.
Thus $lt$ and $rp$ are $\LP_{i-1,\lpath(i-1)}$ and $\RP_{i-1,j}$ \paths\ without intersection with sum
$\PS(lt)+\PS(rp)=\PS(lt)+\PS(rt''[i-1,...,ri'-2])+\PS(rt[ri'-1,...])>
\PS(lt)+\PS(rt)$ that contradict to maximum sum of $lt$ and $rt$.

Thus $\PS(rt[i-1,...,ri'-2])=\PS(rt''[i-1,...,ri'-2])$.
Then using Property \ref{swapable_max}.4 and $ri\le ri'-1$ we get $\PS(rt[i,...,ri-1])=\F_{i,rt(i)}-\F_{ri,rt(ri)}$ that proves this case.

It remains to consider case $ri'\le i'$.
Then $i<ri<ri'\le i'$.

Let $rt_1$ is \concat\ of $rt''[i-1,...,i'-1]$ and $rt'[i',...]$.
By Property \ref{swapable_tails} the $rt_1$ is \path.
By Note \ref{LRP_inherits} the $rt_1$ is $\RP$ \path.
Since $lt(k)<rt'(k)\le rt''(k)$ for each $k=i,...,i'-1$ and $lt(i-1)<j=rt''(i-1)$ then
using Note \ref{concat_without_intersection} the $lt$ don't intersects with $rt_1$.

Since $ri<ri'\le i'$ then $rt'[i',...]=rt[i',...]$ then $rt_1$ intersects $rt$ at row $i'$.
Using Property \ref{swapable_max}.3 we get
$\PS(rt''[i-1,...,i'-1])\ge\PS(rt[i-1,...,i'-1])$.

If $\PS(rt''[i-1,...,i'-1])>\PS(rt[i-1,...,i'-1])$ then
$\PS(rt_1)=\PS(rt''[i-1,...,i'-1])+\PS(rt'[i',...])
%=\F_{i-1,j}-\F_{ri,rt''(ri)}+(\F_{ri,rt''(ri)}-\F_{i',rt''(i')})+\PS(rt[i',...])>\\
\ge\PS(rt[i-1,...,i'-1])+\PS(rt[i',...])=\PS(rt)$.
Then $lt$ and $rt_1$ are $\LP_{i-1,lt(i-1)}$ and $\RP_{i-1,j}$ \paths\ without intersection and with sum $\PS(lt)+\PS(rt_1)>\PS(lt)+\PS(rt)$ that contradict to maximum sum of $lt$ and $rt$.

Then $\PS(rt''[i-1,...,i'-1])=\PS(rt[i-1,...,i'-1])$.

Let $rt_2$ is \concat\ of $rt[i-1,...,i'-1]$ and $rt''[i',...]$.
Since $rt''$ intersects $rt$ at row $i'$ then by Property \ref{swapable_tails}
we get that $rt_2$ is \path.
Then $\PS(rt_2)=\PS(rt[i-1,...,i'-1])+\PS(rt''[i',...])=\PS(rt'')=\F_{i-1,j}$.
Thus using Note \ref{path_defined_by_F_with_sum} we get that $rt_2$ defined by $\F_{i-1,j}$.

Since $ri<i'$ then $rt_2(ri)=rt(ri)$ and $rt_2(i)=rt(i)$.
Then using Property \ref{swapable_max}.1 and $ri<i'$ we get
$\PS(rt[i-1,...,ri])=\PS(rt_2[i-1,...,ri])=\F_{i-1,rt_2(i-1)}-\F_{ri,rt_2(ri)}+\g_{ri,rt(ri)}=\F_{i-1,rt(i-1)}-\F_{ri,rt(ri)}+\g_{ri,rt(ri)}$.
\item

The set $\{b_1,...,b_2\}$ are all possible columns which can be intersected at row $i$ by $\LP_{i-1,\lpath(i-1)}$ \path\ $t_1$ with restriction $t_1(i)<\lpath(i)$.
The set $\{b_3,...,b_4\}$ are all possible columns which can be reached at row $i$ by $\RP_{i-1,j}$ \path\ $t_2$ with restriction $t_2(i)>\rpath(i)$.
Since $\rpath(i)+1\le\W-1$, $\lpath(i-1)\le\lpath(i)$ and $\lpath(i)-1\ge0$ then $b_1\le b_2$ and $b_3\le b_4$
i.e. these sets are not empty.

By Property \ref{nearest_intersection}.3 we get
$\PS(rt[i,...,ri-1])=\F_{i,rt(i)}-\F_{ri,rt(ri)}$.
Since $lt[i,...]$ and $rt'[i,...]$ are $\lrtms(i, lt(i), \rpath(i))$ pair and $lt(i-1)=\lpath(i-1)$
then $\PS(lt[i,...])+\PS(rt'[i,...])=\max_{k=b_1,...,b_2}\{\rMaxes(i,k)\}$.
Recall that $\PS(rt'[i,...,ri-1])=\PS(\rpath[i,...,ri-1])=\F_{i,rt'(i)}-\F_{ri,rt'(ri)}$
and $rt'(ri)=rt(ri)=\rpath(ri)$.
Then\\
$\begin{array}{ll}
\PS(lt[i,...])+\PS(rt[i,...])=
\PS(lt[i,...])+\F_{i,rt(i)}-\F_{ri,rt(ri)}+\PS(rt[ri,...])=\\
=\PS(lt[i,...])+\F_{i,rt(i)}-\F_{i,rt'(i)}+\PS(rt'[i,...,ri-1])+\PS(rt[ri,...])=\\
=\PS(lt[i,...])+\F_{i,rt(i)}-\F_{i,rt'(i)}+\PS(rt'[i,...])=\\
=\PS(lt[i,...])+\PS(rt'[i,...])+\F_{i,rt(i)}-\F_{i,\rpath(i)}.
%=\\
%=\max_{k=b_1,...,b_2}\{\rMaxes(i,k)\}+\F_{i,rt(i)}-\F_{i,\rpath(i)}=\\
%=\max_{k_1\in\{b_1,...,b_2\}, k_2\in\{b_3,...,b_4\}}\big\{\rMaxes(i,k_1)+\F_{i,k_2}\big\}-\F_{i,\rpath(i)}
\end{array}$

Let prove that $\F_{i,rt(i)}=\max_{k=b_3,...,b_4}\{\F_{i,k}\}$.
Since $\rpath(i)<rt(i)$ then $rt(i)\in\{b_3,...,b_4\}$ then $\F_{i,rt(i)}\le\max_{k=b_3,...,b_4}\{\F_{i,k}\}$.

Suppose that exists $j'\in\{b_3,...,b_4\}$ such that $\F_{i,j'}>\F_{i,rt(i)}$.
Consider $\RP_{i,j'}$ \path\ $rt_2$ defined by $\F_{i,j'}$.
Let $rt_2'$ is concatenation of $rt[i-1]$ and $rt_2[i,...]$.
Due to $\max\{j-1,\rpath(i)\}\le b_3\le b_4\le\min\{j+1,\W-1\}$ then $rt_2'$ is $\RP_{i-1,j}$ \path.

If $rt'(k)\le rt_2(k)$ for each $k\ge i$ then $lt$ and $rt_2'$ are $\lrtms(i-1,\lpath(i-1),j)$ pair with sum\\
$\PS(lt)+\PS(rt_2')=\PS(lt)+\F_{i,j'}+\g_{i-1,j}>\PS(lt)+\F_{i,rt(i)}+\g_{i-1,j}\ge\PS(lt)+\PS(rt)$.
That contradict to maximum sum of $lt$ and $rt$.

Then let $i_2\ge i$ such that $rt'(i_2)>rt_2'(i_2)$ and $rt'(k)\le rt_2'(k)$ for each $k=i{-}1,...,i_2{-}1$.
Consider \concat\ $rt_2''$ of $rt_2'[i-1,...,i_2-1]$ and $rt'[i_2,...]$.
By Property \ref{swapable_tails} the $rt_2''$ is \path.
By Note \ref{LRP_inherits} the $rt_2''$ is $\RP_{i-1,j}$ \path.
By Note \ref{concat_without_intersection} the $rt_2''$ don't intersects with $lt$.

If $i_2\ge ri$ then $rt_2''(i_2)=rt'(i_2)=rt(i_2)$ then using Property \ref{swapable_max}.3\\
$\PS(lt)+\PS(rt_2'')=\PS(lt)+\F_{i,j'}-\F_{i_2,rt_2''(i_2)}+\PS(rt[i_2,...])+\g_{i-1,j}>\\
>\PS(lt)+\F_{i,rt(i)}-\F_{i_2,rt(i_2)}+\PS(rt[i_2,...])+\g_{i-1,j}\ge
\PS(lt)+\PS(rt[i,...])+\g_{i-1,j}=\\
=\PS(lt)+\PS(rt)$.
That contradict to maximum sum of $lt$ and $rt$.

Then $i_2<ri$.
Due to $\PS(rt'[i_2,...,ri-1])=\PS(\rpath[i_2,...,ri-1])=\F_{i_2,\rpath(i_2)}-\F_{ri,\rpath(ri)}=\F_{i_2,\rpath(i_2)}-\F_{ri,rt(ri)}$ %and Property \ref{swapable_max}.5
we get\\
$
\PS(lt)+\PS(rt_2'')=
\PS(lt)+\PS(rt_2'[i-1,...,i_2-1])+\PS(rt'[i_2,...,ri-1])+\PS(rt'[ri,...])=\\
=\PS(lt)+\g_{i-1,j}+\F_{i,j'}-\F_{i_2,\rpath(i_2)}+\PS(\rpath[i_2,...,ri-1])+\PS(rt[ri,...])>\\
>\PS(lt)+\g_{i-1,j}+\F_{i,rt(i)}-\F_{ri,rt(ri)}+\PS(rt[ri,...])=\\
=\PS(lt)+\g_{i-1,j}+\PS(rt[i,...,ri-1])+\PS(rt[ri,...])=
\PS(lt)+\g_{i-1,j}+\PS(rt[i,...])=\\
=\PS(lt)+\PS(rt)
$.\\
That contradict to maximum sum of $lt$ and $rt$.

Thus $\F_{i,rt(i)}=\max_{k=b_3,...,b_4}\{\F_{i,k}\}$.
Then
$
\PS(lt[i,...])+\PS(rt'[i,...])+\F_{i,rt(i)}-\F_{i,\rpath(i)}=\\
=\max_{k=b_1,...,b_2}\{\rMaxes(i,k)\}+\max_{k=b_3,...,b_4}\{\F_{i,k}\}-\F_{i,\rpath(i)}.
$
\end{enumerate}
\end {proof}
\end{Prop}

Property \ref{nearest_intersection} divided into 4 key notes, where each next note depends on previous.
%Property \ref{nearest_intersection}.1 can be easy proved using Property \ref{between_intersections}.
%Property \ref{nearest_intersection}.2 can be proved using Property \ref{tail_of_lrtms_is_lrtms}.
Property \ref{nearest_intersection}.3 tells that $rt[i-1,...,ri]$ is subpath of some path defined by $\F_{i-1,rt(i-1)}$.

Property \ref{nearest_intersection}.4 tells that, if we will swap the tails of
$rt[i,...]$ and $\rpath[i,...]$ at
row $ri$, then we get $rt'$ (with head of $\rpath$ and tail of $rt$) for which the difference between $\PS(rt'[i,...])$ and $\PS(rt[i,...])$ is the difference
between $\F_{i,\rpath(i)}$ and $\F_{i, rt(i)}$.

Property \ref{nearest_intersection}.4 is main property that allows to find $\lMaxes$
and $\rMaxes$ in linear time, using DPM.

\begin{Prop} \label{low_bnd}
Let $lt$ and $rt$ are $\lrtms(i-1,\lpath(i-1),j)$ pair.
And let $\max\{0,\lpath(i-1)-1\}\le b_1\le b_2\le\min\{\lpath(i-1)+1,\lpath(i),\rpath(i)-1\}$ and\\
$\max\{\lpath(i)+1,\rpath(i),j-1\}\le b_3\le b_4\le\min\{j+1,\W-1\}$.
Then
$$\PS(lt[i,...])+\PS(rt[i,...])\ge\max_{k=b_1,...,b_2}\{\rMaxes(i,k)\}+\max_{k=b_3,...,b_4}\{\F_{i,k}\}-\F_{i,\rpath(i)}.$$
\begin{proof}
Denote $lt[i,...]$ as $lt^-$ and $rt[i,...]$ as $rt^-$.
And suppose that $\PS(lt^-)+\PS(rt^-)<\rMaxes(i,k_1)+\F_{i,k_2}-\F_{i,\rpath(i)}$
for some $k_1\in\{b_1,...,b_2\}$ and $k_2\in\{b_3,...,b_4\}$.

Let $lt'$ and $rt'$ are $\lrtms(i,k_1,\rpath(i))$ pair.
Then $\PS(lt')+\PS(rt')=\rMaxes(i,k_1)$.

The set $\{b_1,...,b_2\}$ is set of all columns which can be reached by $\LP_{i-1,\lpath(i-1)}$ \path\ at row $i$ except column $\rpath(i)$.
Then concatenation $lt'^+$ of $lt[i-1]$ and $lt'$ is $\LP_{i-1,lt(i-1)}$ \path.
Then by definition of $\F_{i-1,\lpath(i-1)}$ and $\lpath$ we get $\F_{i,k_1}\le\F_{i,\lpath(i)}$.
Since $rt'(i)=\rpath(i)$
then concatenation $rt'^+$ of $\rpath[i-1]$ and $rt'$ is $\RP_{i-1,\rpath(i-1)}$ \path.

Consider $\RP_{i,k_2}$ \path\ $rt''$ defined by $\F_{i,k_2}(\rg)=\F_{i,k_2}$.
The set $\{b_3,...,b_4\}$ is set of all columns which can be reached by $\RP(i-1,j)$ \path\ at row $i$ except column $\lpath(i)$.
Then concatenation $rt''^+$ of $rt[i-1]$ and $rt''$ is $\RP_{i-1,j}$ \path.
By definition $\PS(lt)+\PS(rt)=\lMaxes(i-1,j)\ge\PS(lt'^+)+\PS(rt''^+)$ then $\PS(lt^-)+\PS(rt^-)\ge\PS(lt')+\PS(rt'')$.

\vspace{4px}
If $\rpath(k)<rt''(k)$ for each $k\ge i$ then $\lpath[i-1,...]$ don't intersect $rt''^+$ due to $\lpath(i-1)<j=rt''^+(i-1)$.
Then by definition\\
$\begin{array}{ll}
\PS(lt)+\PS(rt)=\lMaxes(i-1,j)\ge\PS(\lpath[i-1,...])+\PS(rt''^+)=\F_{i-1,\lpath(i-1)}+\g_{i-1,j}+\F_{i,k_2}.
\end{array}$\\
Then $\PS(lt^-)+\PS(rt^-)\ge\F_{i,\lpath(i)}+\F_{i,k_2}\ge\F_{i,k_1}+\F_{i,k_2}\ge\PS(lt')+\F_{i,k_2}$.
Since $rt'^+(i-1)=\rpath(i-1)$ then $\PS(rt'^+)\le\F_{i-1,\rpath(i-1)}$ then $\PS(rt')\le\F_{i,\rpath(i)}$.
Then\\
$\begin{array}{ll}
\PS(lt')+\F_{i,k_2}\ge\PS(lt')+\F_{i,k_2}+\PS(rt')-\F_{i,\rpath(i)}=\rMaxes(i,k_1)+\F_{i,k_2}-\F_{i,\rpath(i)}
\end{array}$.

Thus $\PS(lt^-)+\PS(rt^-)\ge\rMaxes(i,k_1)+\F_{i,k_2}-\F_{i,\rpath(i)}$.
That contradicts to our assumption.

\vspace{4px}
Then exists $ri\ge i-1$ such that $rt''(ri)=\rpath(ri)$.
Then, due to $\rpath(i)=rt'(i)$,
exists $i'\in\{i,...,ri\}$ such that
$rt'(i')\ge rt''(i')$.
WLOG we can assume that $rt'(k)<rt''(k)$ for each $k=i,...,i'-1$ when $i'>i$.
Then $lt'$ don't intersect $rt''[...,i'-1]$.
If $i'=i$ then assume that $rt''[...,i'-1]$ and $rt'[...,i'-1]$ is empty \paths.

Consider \concat\ $rt_1$ of $rt''^+[...,i'-1]$ and $rt'[i',...]$.
By Property \ref{swapable_tails} the $rt_1$ is \path.
By Note \ref{LRP_inherits} the $rt_1[i,...]$ is $\RP_{i,k_2}$ \path.
By Note \ref{concat_without_intersection} $lt'^+[i,...]$ don't intersect $rt_1[i,...]$.

Let $rt_2$ is concatenation of $rt'^+[...,i'-1]$ and $rt''[i',...]$.
By Property \ref{swapable_tails} the $rt_2$ is \path.
Using Prperty \ref{l_max}.1 we get $\PS(rt_2[i,...])\le\PS(\rpath[i,...])=\F_{i,\rpath(i)}$.
Then\\
$\begin{array}{ll}
\PS(rt_1[i,...])&=\PS(rt')+\PS(rt'')-\PS(rt_2[i,...])=
\PS(rt')+\F_{i,k_2}-\PS(rt_2[i,...])\ge\\
&\ge\PS(rt')+\F_{i,k_2}-\F_{i,\rpath(i)}.
\end{array}$

Thus $lt'^+[i,...]$ and $rt_1[i,...]$ are $\LP_{i,k_1}$ and $\RP_{i,k_2}$ \paths\ without intersections and with sum\\
$\begin{array}{ll}
\PS(lt'^+[i,...])+\PS(rt_1[i,...])&\ge
\PS(lt')+\PS(rt')+\F_{i,k_2}-\F_{i,\rpath(i)}=\rMaxes(i,k_1)+\F_{i,k_2}-\F_{i,\rpath(i)}>\\
&>\PS(lt^-)+\PS(rt^-)=\lMaxes(i-1,j)-\g_{i-1,\lpath(i-1)}-\g_{i-1,j}\ge\\
&\ge\PS(lt'^+[i,...])+\PS(rt_1^+[i,...]).
\end{array}$\\
This contradiction proves our Property.
\end{proof}
\end{Prop}

\begin{Lem} \label{_Maxes_is_linear}
Tables $\lMaxes$ and $\rMaxes$ can be found in $\O(\H{\cdot}\W)$.
\begin{proof}
Before calculation of $\lMaxes$ and $\rMaxes$ we need to find table $\F_{i,j}(\g)$ for each $i,j$.
This table can be found in $\O(\H{\cdot}\W)$.
Also, we need in $\lpath$ and $\rpath$, which can be found in $\O(\H)$.

It is enough to prove that every row of tables $\lMaxes$ and $\rMaxes$ can be found in $\O(\W)$.
Let prove it by induction on $\H$.

\vspace{4px}
{\bfseries Base case}:
Let find values for last row.
For last row these tables contains the sum of pair \paths\ with length 1.
Thus, any pair (with different begining) don't intersects between themselves.

\vspace{4px}
$\begin{array}{lll}
\lMaxes(\H{-}1,j)=\g_{\H{-}1, \lpath(\H{-}1)}+\g_{\H{-}1, j}
&$ for each $j=\max\{\rpath(\H{-}1),\lpath(\H{-}1){+}1\}, .., \W{-}1.\\
\rMaxes(\H{-}1,j)=\g_{\H{-}1, \rpath(\H{-}1)}+\g_{\H{-}1, j}
&$ for each $j=0,...,\min\{\lpath(\H{-}1), \rpath(\H{-}1)-1\}.
\end{array}$

\vspace{4px}
This calculation requires $\O(\W)$ time.

\vspace{4px}
{\bfseries Induction step}:
Suppose that known $\lMaxes$ and $\rMaxes$ for rows $i,...,{\H{-}1}$, where $i>0$.

Then let find $\lMaxes$ for $(i-1)$-th row.
By Note \ref{M_defined} it is enough to find the $\lMaxes(i-1,j)$, for any $j\ge\max\{\lpath(i{-}1)+1,\rpath(i{-}1)\}$.

Let $lP$ and $rP$ are $\lrtms(i-1,\lpath(i-1),j)$ pair.
Consider all possible cases and find the sum $\PS(lP[i,...])+\PS(rP[i,...])$:
\begin{enumerate}
\item
For case $lP(i)<\lpath(i)$ and $rP(i)=\rpath(i)$.
Denote $\PS(lP[i,...])+\PS(rP[i,...])$ for this case as $max_1(j)$.
Then we get $max_1(j)=\rMaxes(i,lP(i))$ i.e.
$$
	 max_1(j)=\max_{k=b_1,...,b_2}\{\rMaxes(i,k)\}
$$
where $b_1=\max\{\lpath(i-1)-1,0\}$ and
$b_2=\min\{\lpath(i-1)+1,\lpath(i)-1,\rpath(i)-1\}$

Due to rule (r1) we get $\lpath(i-1)+1\ge\lpath(i)$ and $\lpath(i)-1\le\rpath(i)-1$,
then $b_2=\lpath(i)-1$.
Note that $b_1\le b_2$ iif $\max\{\lpath(i{-}1),1\}\le\lpath(i)$.

Let find when restrictions of this case don't contradict to (r1), (r3).
It is enough to check for possible positions of $lP(i)$ and $rP(i)$.

For $rP(i)$ we get $j{-}1\le rP(i)\le j{+}1$ and $rP(i)=\rpath(i)<\W$,
then sufficient conditions for $rP(i)$ are $j{-}1\le\rpath(i)\le j{+}1$.
But by proposition $j\ge\rpath(i-1)$ then by (r1) the condition $j\ge\rpath(i)-1$ is true always.

Restrictions for $lP(i)$ are $\lpath(i-1)-1=lP(i-1)-1\le lP(i)<\lpath(i)$ and $0\le lP(i)$.

Thus we get conditions when this case need to check 
\begin{equation}\label{cond1}
j{-}1\le\rpath(i), \max\{\lpath(i{-}1), 1\}\le\lpath(i).
\end{equation}
Thus, in common case, we can assume
$$
max_1(j)=\begin{cases}
\max_{k=b_1,...,b_2}\{\rMaxes(i,k)\} &(\ref{cond1}),\\
0 &otherwise.
\end{cases}
$$
\item
For case $lP(i)=\lpath(i)$.
Denote $\PS(lP[i,...])+\PS(rP[i,...])$ for this case as $max_2(j)$.
Then we get $max_2(j)=\lMaxes(i,rP(i))$ i.e.
$$
	 max_2(j)=\max_{k=b_1,...,b_2}\{\lMaxes(i,k)\}
$$
where $b_1=\max\{j-1,\rpath(i),\lpath(i)+1\}$ and
$b_2=\min\{j+1,\W-1\}$

Note that $b_1\le b_2$ iif $lpath(i)+2\le\W$.

For $rP(i)$ we get restrictions $\max\{\lpath(i)+1,\rpath(i)\}\le rP(i)\le\W-1$
and $j-1\le rP(i)\le j+1$.
Since always $\rpath(i)\le\min\{j+1,\W-1\}$ and $j-1\le\W-1$ then required conditions
for $rP(i)$ are $lpath(i)+1\le j+1$ and $\lpath(i)+2\le\W$.
%I.e. $\lpath(i)\le j$.
But by proposition and (r1) we get $j\ge\lpath(i-1)+1\ge\lpath(i)$ then we get that $\lpath(i)\le j$ is true always.

Restrictions for $lP(i)$ are $lP(i)=\lpath(i)$ and $lP(i-1)=\lpath(i-1)$.
Since $\lpath[i-1,i]$ satisfy to (r1) and (r3) then this restriction always true for $lP[i-1,i]$.

Thus we get conditions for this case checking 
\begin{equation}\label{cond2}
\lpath(i)+2\le\W.
\end{equation}
Thus, in common case, we can assume
$$
max_2(j)=\begin{cases}
\max_{k=b_1,...,b_2}\{\lMaxes(i,k)\} &(\ref{cond2}),\\
0 &otherwise.
\end{cases}
$$
\item
Consider case when $lP(i)<\lpath(i)$, $rP(i)>\rpath(i)$ and $j=\rpath(i-1)$.

Due to contradiction with Properties \ref{between_intersections}.1 and \ref{between_intersections}.2 this case impossible for $\lrtms(i-1,\lpath(i-1),j)$ pair $lP$ and $rP$.
\item
Consider case when $lP(i)<\lpath(i)$, $rP(i)>\rpath(i)$ and $j>\rpath(i-1)$.
Denote $\PS(lP[i,...])+\PS(rP[i,...])$ for this case as $max_3(j)$.
Then by Property \ref{nearest_intersection}.4 we get
$$max_3(j)=\max_{k=b_1,...,b_2}\{\rMaxes(i,k)\}+\max_{k=b_3,...,b_4}\{\F_{i,k}\}-\F_{i,\rpath(i)}$$
where $b_1=\max\{0,\lpath(i-1)-1\}, b_2=\min\{\lpath(i-1)+1,\lpath(i)-1\}=\lpath(i)-1$ and\\
$b_3=\max\{\rpath(i)+1,j-1\}, b_4=\min\{j+1,\W-1\}$.

Note that $b_1\le b_2$ and $b_3\le b_4$ iif $\max\{1,\lpath(i-1)\}\le\lpath(i),\ \rpath(i)+2\le\W$.

This case possible only when $\rpath(i)<rP(i)\le\W-1$, $j{-}1\le rP(i)\le j{+}1$,
$\rpath(i{-}1)<j$, $\lpath(i{-}1){-}1\le lP(i)<\lpath(i)$ and $0\le lP(i)$.
Then we get condition of $max_3(j)$ existing
\begin{equation}\label{cond3}
\max\{1,\lpath(i{-}1)\}\le\lpath(i),\ \rpath(i){+}2\le\W,\ \rpath(i{-}1)<j.
\end{equation}
Thus, in common case, we can assume
$$
max_3(j)=\begin{cases}
\max_{k=b_1,...,b_2}\{\rMaxes(i,k)\}+\max_{k=b_3,...,b_4}\{\F_{i,k}\}-\F_{i,\rpath(i)} &(\ref{cond3}),\\
0 &otherwise.
\end{cases}
$$
Note that condition $b_1\le b_2$ and $b_3\le b_4$ follows from $(\ref{cond3})$.
\end{enumerate}

Thus exists $m\in\{1,2,3\}$ such that $\PS(lP[i,...])+\PS(rP[i,...])=max_m(j)$.
Then
$$
\PS(lP[i,...])+\PS(rP[i,...])\le\max\{max_1(j),max_2(j),max_3(j)\}.
$$

Since $\PS(lP[i,...])+\PS(rP[i,...])\ge0$ then $\PS(lP[i,...])+\PS(rP[i,...])\ge max_m(j)$
when condition $(m)$ is false for each $m\in\{1,2,3\}$.
Since $max_1(j)$ and $max_2(j)$ is result of reducing to an existing pairs of \paths\ with
maximum sum then $\PS(lP[i,...])+\PS(rP[i,...])\ge max_m(j)$ for each $m\in\{1,2\}$.

Since $b_1\le b_2$ and $b_3\le b_4$ in case 4 follows from condition $(\ref{cond3})$ then
by Propety \ref{low_bnd} we get that $\PS(lP[i,...])+\PS(rP[i,...])\ge max_3(j)$.
Thus using $\lMaxes(i-1,j)=\PS(lP[i,...])+\PS(rP[i,...])+\g_{i-1,\lpath(i-1)}+\g_{i-1,j}$
we get
$$
\lMaxes(i-1,j)=\g_{i-1,\lpath(i-1)}+\g_{i-1,j}+\max\{max_1(j),max_2(j),max_3(j)\}.
$$

Thus in $\O(1)$ we can find $\lMaxes(i-1,j)$ for any $j\in\{\max\{\lpath(i{-}1)+1,\rpath(i{-}1)\},...,\W-1\}$.
Then in $\O(\W)$ we can find $\lMaxes$ for row $i-1$.
Similarly in $\O(\W)$ we can find $\rMaxes$ for row $i-1$.
\end{proof}
\end{Lem}

More exactly, this algorithm spent $\O(\H{\cdot}\W)$ of comparisons and sums of
numbers like $\F_{i,j}$, $\lMaxes(i,j)$, $\lpath(i)$.
Since values of $\g$ bounded by value $C$, then these numbers have
length $\O(\log(\H\cdot C))$.

If $C=f(H)$ is less than any polynomial function of $\H$,
then these values have length $o(\log(\H))$ i.e. less than length
of addresses to elements of input data, therefore we ignore linear operations with these values.

If $f$ is not less than some polynomial function of $\H$,
then complexity is $\O(\H{\cdot}\W)$ of linear operations with integers of length
$\O(\log(f(\H)))$.
Thus, we have full complexity $\O(\H{\cdot}\W{\cdot}\log(f(\H)))$.
But by our assumption, the length of input data is $\Theta(\H{\cdot}\W{\cdot}\log(f(\H)))$.
Thus, we got pure linear algorithm.

\subsubsection{Simplification of $\lMaxes$ and $\rMaxes$ search}

Here we use designations from induction step of Lemma \ref{_Maxes_is_linear}.

Assume that $lP(i)<\lpath(i)$.
Note that pair $b_1,b_2$ of case 1 are same as pair $b_1,b_2$ of case 4.
Also using restriction $rP(i)=\rpath(i)$ in case 1 we get
$$
\max_{k=\rpath(i),...,b_4}\{\F_{i,k}\}=\F_{i,\rpath(i)}
$$
for any $\rpath(i)\le b_4\le\min\{j+1,\W-1\}$.
Thus we can assume that $b_4$ from case 4 and
$$
max_1(j)=\max_{k=b_1,...,b_2}\{\rMaxes(i,k)\}
+\max_{k=\rpath(i),...,b_4}\{\F_{i,k}\}-\F_{i,\rpath(i)}.
$$
Also we can extend restriction for case 4 by addition of restriction of cases 1 and 3.
Let $b_3'=\max\{\rpath(i), j-1\}$.
Then in case $rP(i)=\rpath(i)$ %and $j\ge\max\{\lpath(i-1)+1, \rpath(i-1)\}$
we get $\rpath(i)=rP(i)\ge j-1$ then
we get $b_3'=\rpath(i)$ then
$$
max_1(j)=\max_{k=b_1,...,b_2}\{\rMaxes(i,k)\}
+\max_{k=b_3',...,b_4}\{\F_{i,k}\}-\F_{i,\rpath(i)}.
$$
If $rP(i)>\rpath(i)$ then by case 3 we get that case $j=\rpath(i-1)$ impossible.
Then $j>\rpath(i-1)$ and we get restrictions of case 4 and conditions of Property \ref{nearest_intersection}.

In case when $j{-}1>\rpath(i)$ we get $b_3'=b_3$.

Consider case when $j{-}1\le\rpath(i)$ i.e. $b_3'=\rpath(i)=b_3-1$.
Then by Property \ref{nearest_intersection} exists $ri>i$ such that $rP[i-1,...,ri]$ is
\subpath\ of some $\RP$ \path\ defined by $\F$ then
$$
rP(i)\in\argmax_{k=b_3',...,b_4}\{\F_{i,k}\}.
$$
Since $b_3'=\rpath(i)<rP(i)$ and $b_3'+1=b_3$ then $rP(i)\in\{b_3,...,b_4\}$ then
$$
\max_{k=b_3,...,b_4}\{\F_{i,k}\}=\max_{k=b_3',...,b_4}\{\F_{i,k}\}%\ge\F_{i,\rpath(i)}
$$
Thus if $rP(i)>\rpath(i)$ we get
$$
max_3(j)=\max_{k=b_1,...,b_2}\{\rMaxes(i,k)\}
+\max_{k=b_3',...,b_4}\{\F_{i,k}\}-\F_{i,\rpath(i)}.
$$

Thus, in common case, we can combine cases 1, 3 and 4 with one restriction $lP(i)<\lpath(i)$
and common maximum formula
$$
max_1'(j)=\max_{k=b_1,...,b_2}\{\rMaxes(i,k)\}
+\max_{k=b_3',...,b_4}\{\F_{i,k}\}-\F_{i,\rpath(i)}.
$$
Let find conditions of $max_1'$ existing.

For $rP(i)$ we get $\rpath(i)\le rP(i)\le\W{-}1$ and $j{-}1\le rP(i)\le j{+}1$ then we get $\rpath(i)\le j+1$.
But $j=rP(i-1)\ge rP(i)-1\ge\rpath(i)+1$ allways.

For $lP(i)$ we get $\lpath(i{-}1){-}1\le lP(i)<\lpath(i)$ and $0\le lP(i)$.
Then we get conditions of $max_1'(j)$ existing
\begin{equation}\label{cond4}
\max\{1,\lpath(i{-}1)\}\le\lpath(i).
\end{equation}
Thus in common case we can assume
$$
max_1'(j)=\begin{cases}
\max_{k=b_1,...,b_2}\{\rMaxes(i,k)\}+\max_{k=b_3',...,b_4}\{\F_{i,k}\}-\F_{i,\rpath(i)} &(\ref{cond4})\\
0 &otherwise.
\end{cases}
$$
Then
$$
\lMaxes(i-1,j)=\g_{i-1,\lpath(i-1)}+\g_{i-1,j}+\max\{max_1'(j),max_2(j)\}
$$

Implementation of this version search of $\lMaxes$ and $\rMaxes$ represented at listing 3 in function get\_M\ using programing language Python.

\subsection{Reducing problem to \lrtms(0,0,\W-1) pair}

\begin{Def}
Denote subset of common cells between \paths\ $t_1$ and $t_2$ as $t_1\cap t_2$.\\
Set of all cells of \paths\ $t_1$ and $t_2$ as $t_1\cup t_2$.\\
Set of all cells of \path\ $t_1$ without cells of \path\ $t_2$ as $t_1\setminus t_2$.
\end{Def}

\begin{Def}
Consider \paths\ $lt$ and $rt$.
Let rows $i_1$ and $i_2$ such that $lt(i)=rt(i)$ for each $i=i_1+1,...,i_2-1$ and either
$lt(i_1)<rt(i_1),\ lt(i_2)>rt(i_2)$ or $lt(i_1)>rt(i_1),\ lt(i_2)<rt(i_2)$.
Then call pair $i_1,i_2$ as {\textbf {cross over pair}}.
\end{Def}

\begin{Prop} \label{cross_over_off}
For any \paths\ $lt$ and $rt$, with begining from cells $(0,0)$ and $(0,\W-1)$,
exists \paths\ $lt'$ and $rt'$ with begining from $(0,0)$ and $(0,\W-1)$ respectively,
with $lt\cup rt=lt'\cup rt'$ (as corrolary with same common sum i.e. $\PS(lt\cup rt)=\PS(lt'\cup rt')$), and
inequality $lt'(i)\le rt'(i)$ for each $i$.
\begin{proof}
%Denote $\PS(lt)+\PS(rt)-\PS(lt\cap rt)$ as $N$.
WLOG suppose that $lt$ and $rt$ have minimum cross over pairs from all \paths\ $lt'$ and $rt'$ starts from $(0,0)$ and $(0,\W-1)$ respectively with same common sum (equal to $N$),
and $lt\cup rt=lt'\cup rt'$.
And suppose that between $lt$ and $rt$ exists cross over pair.

%Let find $lmp_1$ and $rmp_1$ such that: $lmp_1(i)\le rmp_1(i)$ for each $i$, and $\PS(lmp_1)+\PS(rmp_1)-\PS(lmp_1\cap rmp_1)=N$.

Then, using Property \ref{swapable_tails}, we can reduce number of cross over pairs
by swaping tails of $lt$ and $rt$.
Since swaping don't changes the set of cells of \paths\ then we get $lt\cup rt=lt'\cup rt'$.
Thus we get contradiction with minimum cross over pairs between $lt$ and $rt$.

%Consider $i_1$-th row such that $lmp(i)<rmp(i)$ for each $i\le i_1$ and $lmp(i_1)>rmp(i_1)$. Then by Property \ref{swapable_tails} tails of $lmp$ and $rmp$ are swapable from $i$-th row. Then let $lmp'$ and $rmp'$ are \paths\ after swaping these tails. I.e.:\\$\begin{array}{llll}lmp'[...,i_1-1]&=lmp[...,i_1-1], &lmp'[i_1,...]&=lmp[i_1,...]\\rmp'[...,i_1-1]&=rmp[...,i_1-1], &rmp'[i_1,...]&=rmp[i_1,...]\end{array}$

%Thus we found $lmp'$ and $rmp'$ \paths\ whith $\PS(lmp')+\PS(rmp')-\PS(lmp'\cap rmp')=N$. Also, if exists $i$ such that $lmp'(i)>rmp'(i)$, then $i>i_1$.
%Thus we can reduce the length of tails which have these rows.

Thus we get $lt(i)\le rt(i)$ for each $i$.
\end{proof}
\end{Prop}

\begin{Prop} \label{intersection_off}
Suppose that our grid $g$ without negative values.
Consider \paths\ $lt$ and $rt$ with begining from $(0,0)$ and $(0,\W-1)$, and
$lt(i)\le rt(i)$ for each $i$.

Then exists \paths\ $lt'$ and $rt'$ with begining from $(0,0)$ and $(0,\W-1)$ respectively,
such that $lt'(i)<rt'(i)$ for each $i$
(i.e. $lt'$ don't intersects with $rt'$),
and $\PS(lt')+\PS(rt')\ge\PS(lt)+\PS(rt)-\PS(lt\cap rt)$.

\begin{proof}
Denote $\PS(lt)+\PS(rt)-\PS(lt\cap rt)$ as $N$.
WLOG assume that $lt$ and $rt$ have minimum common cells among all \paths\ starts from $(0,0)$ and $(0,W-1)$ cells, and with common sum equal to $N$ or grater (i.e. $\PS(lt)+\PS(rt)-\PS(lt\cap rt)\ge N$).

%Using $lmp_1$ and $rmp_1$ we can find $lmp'$ and $rmp'$ without intersections and $\PS(lmp')+\PS(rmp')=N$.

And suppose that row $i_1$ such that $lt(i_1)=rt(i_1)$ and $lt(i)<rt(i)$ for each $i<i_1$.
Denote $lt(i_1)$ as $j_1$.

Consider case when $lt(i_1-1)<j_1$.

Due to rule of moving (r1), after $k$ steps from cell $(i_1,j_1)$ left and rigth
robots will be located on cells $(i_1+k,j')$ and $(i_1+k,j'')$ respectively,
for some $j',j''\le j_1+k$.
I.e. $lt(i_1+k)\le rt(i_1+k)\le j_1+k$.

Consider cases:
\begin{itemize}%[topsep=0pt, partopsep=0pt]%[noitemsep,topsep=0pt,parsep=0pt,partopsep=0pt]
\item
Suppose that not all moves of left robot are rigthmost after row $i_1$.
I.e. exists $i'>i_1$ such that ${(lt(i)-j_1)\ge(i{-}i_1)}$ for each $i=i_1,...,i'{-}1$
and $(lt(i')-j_1)<(i'{-}i_1)$,

Then
$j_1+i-i_1\le lt(i_1+i-i_1)\le rt(i_1+i-i_1)\le j_1+i-i_1$ for each $i=i_1,...,i'{-}1$.
I.e. $lt[i_1,...,i'-1]=rt[i_1,...,i'-1]$.

Consider \concat\ $lmp'$ such that:\\
$\begin{array}{lll}
lmp'[...,i_1-1]&=lt[...,i_1-1],\\
lmp'(i_1+k)&=j_1-1+k,\ k=0,...,i'-1-i_1,\\
lmp'[i',...]&=lt[i',...].
\end{array}$

Then $lmp'[...,i_1-1]$ and $lmp'[i',...]$ are \subpaths.
Also, $lmp'[i_1,...,i'-1]$ is \subpath\ with rigthmost moves.

\vspace{4px}
Let prove that moves from $lmp'(i_1-1)$ to $lmp'(i_1)$ and from $lmp'(i'-1)$ to $lmp'(i')$
are corresponds to move rules.

Using rules of move for $lt$ we get
$lmp'(i_1-1)=lt(i_1-1)\ge lt(i_1)-1=lmp'(i_1)$.
The other side $lmp'(i_1)=lt(i_1)-1>lt(i_1-1)-1=lmp'(i_1-1)-1$.
I.e. $lmp'(i_1)=lmp'(i_1-1)$.
Thus move from $lmp'(i_1-1)$ to $lmp'(i_1)$ is correct (i.e. corresponds to moving rules).

\vspace{4px}
By assumption $lmp(i')-j_1<(i'-i_1)$ we get $j_1>lt(i')-(i'-i_1)$.
Then for $k=i'-1-i_1$ we get $lmp'(i'-1)=lmp'(i_1+k)=j_1-1+k>lt(i')-2=lmp'(i')-2$.
I.e. $lmp'(i'-1)\ge lmp'(i')-1$.

By assumption $lt(i'-1)-j_1\ge (i'-1-i_1)$ we get $j_1\le lt(i'-1)-(i'-1-i_1)$.
Then for $k=i'-1-i_1$ we get $lmp'(i'-1)=j_1+k-1\le lt(i'-1)-1\le lt(i')$.

I.e. we get $lmp'(i'-1)\le lmp'(i')\le lmp'(i'-1)+1$.
Then move from $lmp'(i'-1)$ to $lmp'(i')$ is correct too.
Thus $lmp'$ is \path.

\vspace{4px}
By definition $lmp'(i)=j_1-1+(i-i_1)$ for each $i=i_1,...,i'-1$.
Then, using assumption $lt(i)-j_1\ge (i-i_1)$ for each $i=i_1,...,i'-1$,
we get $(lt(i)-j_1+(j_1-1+(i-i_1)))\ge(i-i_1)+lmp'(i)$ for each $i=i_1,...,i'-1$.
I.e. $lt(i)\ne lmp'(i),i=i_1,...,i'-1$.

\vspace{4px}
Denote $\PS(lmp_1\cap rmp_1)$ and $\PS(lmp'\cap rt)$ as $d$ and $d'$ respectively.

Since $lmp'(i)\ne lt(i)=rt(i)$ for each $i=i_1,...,i'{-}1$, then
$d'=d-\PS(lt[i_1,...,i'{-}1])$.
Since $g$ consists of nonnegative values, then $\PS(lmp'[i_1,...,i'-1])\ge0$.
Then\\
$\begin{array}{ll}
N&=\PS(lt)+\PS(rt)-d=\\
&=\PS(lt[...,i_1{-}1])+\PS(lt[i_1,...,i'{-}1])+\PS(lt[i',...])+\PS(rt)-d=\\
&=\PS(lt[...,i_1{-}1])+\PS(lt[i',...])+\PS(rmp)-d'\le\\
&\le\PS(lt[...,i_1{-}1])+\PS(lmp'[i_1,...,i'{-}1])+\PS(lt[i',...])+\PS(rt)-d'=\\
&=\PS(lmp')+\PS(rt)-\PS(lmp'\cap rt).
\end{array}$

\vspace{4px}
Thus we get \paths\ $lmp'$ and $rt$ from $(0,0)$ and $(0,W{-}1)$ respectively with common sum not less than common sum of $lt$ and $rt$.
Since $rt$ has common cells with $lmp'$ less than with $lt$,
then we get contradiction with minimum of common cells between $lt$ and $rt$.
\item
${(lt(i)-j_1)\ge(i{-}i_1)}$ for each $i\ge i_1$.

Then
$j_1+i-i_1\le lt(i)\le rt(i_1+(i-i_1))\le j_1+(i-i_1)$ for each $i\ge i_1$.
I.e. $lt[i_1,...]=rt[i_1,...]$ and $lt(i)=j_1+i-i_1$ for each $i\ge i_1$.

Consider \concat\ $lmp'$ such that:\\
$\begin{array}{lll}
lmp'[...,i_1-1]&=lt[...,i_1-1],\\
lmp'(i)&=lt(i_1-1)$ for each $i\ge i_1.
\end{array}$

Then $lmp'[...,i_1{-}1]$ and $lmp'[i_1,...]$ are \paths.
Also, $lmp'(i_1)=lt(i_1{-}1)=lmp'(i_1{-}1)$ i.e. move from $lmp'(i_1{-}1)$ to $=lmp'(i_1)$ is correct.
Thus $lmp'$ is \path.

Also, $lmp'(i)=lt(i_1-1)<j_1\le j_1+i-i_1=lt(i)$ for each $i\ge i_1$.

\vspace{4px}
Denote $\PS(lt\cap rt)$ and $\PS(lmp'\cap rt)$ as $d$ and $d'$ respectively.

Since $lmp'(i)<lt(i)=rt(i)$ for each $i\ge i_1$, then $d'=d-\PS(lt[i_1,...])$.
Since $g$ consists of nonnegative values, then $\PS(lmp'[i_1,...,i'-1])\ge0$.
Then\\
$\begin{array}{ll}
N&=\PS(lt)+\PS(rt)-d=
\PS(lt[...,i_1{-}1])+\PS(lt[i_1,...])+\PS(rt)-d=\\
&=\PS(lt[...,i_1{-}1])+\PS(rt)-d'\le\\
&\le\PS(lt[...,i_1{-}1])+\PS(lmp'[i_1,...])+\PS(rt)-d'=\\
&=\PS(lmp')+\PS(rt)-\PS(lmp'\cap rt).
\end{array}$

Thus, like in previous case, we get contradiction with minimum of common cells between $lt$ and $rt$.
\end{itemize}

It remains to consider case when $lt(i_1-1)\ge j_1$.
Then $rt(i_1-1)>lt(i_1-1)\ge j_1=rt(i_1)$, and, due to simmetry, this case lead us to contradiction like in previous case.
\end{proof}
\end{Prop}

\begin{Prop} \label{lr_to_LRP}
Consider \paths\ $lt$ and $rt$ with begining from $(0,0)$ and $(0,\W-1)$ respectively,
and $lt(i)<rt(i)$ for each $i$.
Then exists $\LP(0,0)$ and $\RP(0,\W-1)$ \paths\ $lt'$ and $rt'$ respectively
such that $lt'(i)<rt'(i)$ for each $i$,
and $\PS(lt')+\PS(rt')\ge\PS(lt)+\PS(rt)$.
\begin{proof}
Denote $\PS(lt)+\PS(rt)$ as $N$.
WLOG we can assume that $lt(i)$ has minimum amount of rows $i$
such that $lt(i)>\lpath(i)$ among all \paths\ $lt'$ with begining on $(0,0)$ without intersections with $rt$, and with sum $\PS(lt')+\PS(rt)\ge N$.

Suppose that $lt(i)$ isn't $\LP_{0,0}$ \path.
Then exists row $i_1$ such that $lt(i)\le\lpath(i)$ for each $i<i_1$ and $lt(i_1)>\lpath(i_1)$ (i.e. $i_1>0$).
Then consider cases:
\begin{itemize}
\item
If exists $i_2>i$ such that $lt(i)>\lpath(i)$ for each $i=i_1,...,i_2-1$ and
$lt(i_2)\le\lpath(i_2)$.

Then consider concatenation $t_1$:
$t_1[...,i_1-1]=lt[...,i_1-1]$,
$t_1[i_1,...]=\lpath[i_1,...]$.

And concatenation $lmp'$:
$lmp'[...,i_2-1]=t_1[...,i_2-1]$,
$lmp'[i_2,...]=lt[i_2,...]$.

Due to Property \ref{swapable_tails} the $t_1$ is \path. Then due to Property \ref{swapable_tails} the $lmp'$ is \path\ too.

Thus we get \path\ $lmp'$:\\
$\begin{array}{ll}
lmp'[...,i_1-1]&=lt[...,i_1-1],\\
lmp'[i_1,...,2_1-1]&=\lpath[i_1,...,i_2-1],\\
lmp'[i_2,...]&=lt[i_2,...].
\end{array}$

Sumilarly we can prove that concatenation $t_2$:\\
$\begin{array}{ll}
t_2[...,i_1-1]&=\lpath[...,i_1-1],\\
t_2[i_1,...,2_1-1]&=lt[i_1,...,i_2-1],\\
t_2[i_2,...]&=\lpath[i_2,...].
\end{array}$\\
is \path\ too.

Due to $\lpath$ defined by $\F_{0,0}$ and $t_2$ is path with begining on $(0,0)$
\\%, and using Property \ref{swapable_max}.1 and Note we get \ref{tail_of_max}\\
%$\begin{array}{ll}
%\PS(lmp[i_1,...,i_2-1])&=\PS(t_2)-\PS(\lpath[...,i_1-1])-\PS(\lpath[i_2,...])\le\\
%&\le\F_{0,0}-(\F_{0,0}-\F_{i_1,\lpath(i_1)})-\F_{i_2,\lpath(i_2)}=\\
%&=\F_{i_1,\lpath(i_1)}-\F_{i_2,\lpath(i_2)}
%\end{array}$
$\begin{array}{ll}
\PS(lt[i_1,...,i_2{-}1])&=\PS(t_2)-\PS(\lpath[...,i_1{-}1])-\PS(\lpath[i_2,...])\le\\
&\le\PS(\lpath)-\PS(\lpath[...,i_1{-}1])-\PS(\lpath[i_2,...])=\\
&=\PS(\lpath[i_1,...,i_2{-}1]).
\end{array}$

Then\\
$\begin{array}{ll}
\PS(lmp')&=\PS(lt[...,i_1-1])+\PS(\lpath[i_1,...,i_2-1])+\PS(lt[i_2,...])\ge\\
&\ge\PS(lt[...,i_1-1])+\PS(lt[i_1,...,i_2-1])+\PS(lt[i_2,...])=\PS(lt).
\end{array}$

Since $\lpath(i)\le lt(i)$ for each $i=i_1,...,i_2-1$,
then $lmp'(i)\le lt(i)<rt(i)$ for each $i$.

Thus we get \path\ $lmp'$ without intersections with $rt$
and $\PS(lmp')+\PS(rt)\ge\PS(lt)+\PS(rt)=N$.

But $lmp'$ has less rows $i$ such that $lmp'(i)>\lpath(i)$
which contradicts to minimum of these rows in $lt$.
Thus $lt$ is $\LP_{0,0,}$ \path.

\item
$lt(i)>lpath(i)$ for each $i\ge i_1$.

Then consider concatenations $lmp'$ and $t_2$:\\
$\begin{array}{llll}
lmp'[...,i_1-1]&=lt[...,i_1-1],
&lmp'[i_1,...]&=\lpath[i_1,...],\\
t_2[...,1_1-1]&=lpath[...,i_1-1],
&t_2[i_1,...]&=lt[i_1,...].
\end{array}$

Due to Property \ref{swapable_tails} the $lmp'$ and $t_2$ are \paths.

Due to $\lpath$ defined by $\F_{0,0}$ and $t_2$ is path with begining on $(0,0)$\\
$\begin{array}{ll}
\PS(lt[i_1,...])&=\PS(t_2)-\PS(\lpath[...,i_1{-}1])\le\\
&\le\PS(\lpath)-\PS(\lpath[...,i_1{-}1])=\PS(\lpath[i_1,...]).
\end{array}$

Then\\
$\begin{array}{ll}
\PS(lmp')&=\PS(lt[...,i_1-1])+\PS(\lpath[i_1,...])\ge\\
&\ge\PS(lt[...,i_1-1])+\PS(lt[i_1,...])=\PS(lt).
\end{array}$

Since $\lpath(i)\le lt(i)$ for each $i\ge i_1$,
then $lmp'(i)\le lt(i)<rt(i)$ for each $i$.

Thus we get \path\ $lmp'$ without intersections with $rt$
and $\PS(lmp')+\PS(rt)\ge\PS(lt)+\PS(rt)=N$.

But $lmp'$ has less rows $i$ such that $lmp'(i)>\lpath(i)$
which contradicts to minimum of these rows in $lt$.
Thus $lt$ is $\LP_{0,0,}$ \path.

\end{itemize}

Similarly we can prove that $rt$ is $\RP_{0,W-1}$ \path.
\end{proof}
\end{Prop}

\begin{Def}
Consider pair of \paths\ $l$ and $r$, with intersection in $i$-th row.
Assume that there are no \paths\ $l'$ and $r'$ such that they contains all cells of $l$ and $r$, but without intersections at $i$-th row (i.e. $(l\cup r)\varsubsetneq(l'\cup r')$),
and all cells $(l'\cup r')\setminus(l\cup r)$ with nonnegative values.
Then call \paths\ $l$ and $r$ as $(i,l(i))$--linked pair.
And call cell $(i,j)$ as {\textbf{bottleneck}} if there are $(i,j)$--linked pair.
\end{Def}

\begin{Lem} \label{best_solve_in_lrtmspair}
Let $N$ is maximum number of cherries which can be collected by 2 robots with begining on $(0,0)$ and $(0,\W{-}1)$ cells. If is true at least one of next conditions:

\vspace{-6px}
\begin{enumerate}
\item\label{best_solve_in_lrtmspair:for_nonnegative}
All values of grid $\g$ is nonnegative.

\vspace{-6px}
\item\label{best_solve_in_lrtmspair:3}
The $\g$ don't has bottlenecks.
\end{enumerate}

\vspace{-6px}
then any $\lrtms(0,0,\W{-}1)$ pair $lt$ and $rt$ have $\PS(lt)+\PS(rt)=N$.
\begin{proof}
Let \tracks\ $lmp$ and $rmp$ starts from $(0,0)$ and $(0,\W-1)$ cells respectively and pickups maximum cherries. I.e. $\PS(lmp)+\PS(rmp)-\PS(lmp\cap rmp)=N$.

By Property \ref{cross_over_off} we can assume that $lmp(i)\le rmp(i)$ for each $i$.
\begin{enumerate}
\item
Suppose that all values of $\g$ is nonnegative.

Then by Property \ref{intersection_off} we can assume that $lmp(i)<rmp(i)$ for each $i$.
\item
Suppose that $\g$ don't has bottlenecks.

WLOG assume that $lt$ and $rt$ have minimum of intersections among all pairs of \paths\ with
begining from $(0,0)$ and $(0,\W-1)$, and common sum equal to $N$ or grater.

Suppose that $lt$ intersects with $rt$.

Since grid $\g$ don't has bottlenecks, then exists \paths\ $lt'$ and $rt'$ such that
$(lt\cup rt)\varsubsetneq(lt'\cup rt')$, and cells
$(lt'\cup rt')\setminus(lt\cup rt)$ without negative values.

By Property \ref{cross_over_off} exists $lt''$ and $rt''$ started from $(0,0)$ and $(0,\W-1)$
without cross over pairs, and $(lt'\cup rt')=(lt''\cup rt'')$.

Thus we get \paths\ $lt''$ and $rt''$ started from $(0,0)$ and $(0,\W-1)$ such that $lt''(i)\le rt''(i)$ for each $i$, and with common sum\\
$\PS(lt''\cup rt'')=\PS(lt'\cup rt')=\PS(lt\cup rt)+\PS((lt'\cup rt')\setminus(lt\cup rt))\ge\PS(lt\cup rt)=N$.

But $lt''$ and $rt''$ have less intersections than $lt$ and $rt$,
that contradicts with our assumption.

Thus $lt$ don't intersects with $rt$.
Then $lt(i)<rt(i)$ for each $i$.
Then, as in previous case, we can assume that $lmp(i)<rmp(i)$ for each $i$.
\end{enumerate}

Then by Property \ref{lr_to_LRP} exists $\LP_{0,0}$ and $\RP_{0,W{-}1}$ \paths\ $lmp'$ and $rmp'$ respectively without intersections, and $\PS(lmp')+\PS(rmp')=N$.
Since $N$ is upper bound for collected cherries by any pair of $\LP_{0,0}$ and $\RP_{0,\W{-}1}$ \paths\,
then $lmp'$ and $rmp'$ are $\lrtms(0,0,\W-1)$ pair.

Due to uniqueness of maximum, all $\lrtms(0,0,\W{-}1)$ pairs have same sum i.e. $N$.
\end{proof}
\end{Lem}

\section{Linear solution}

\begin{Teo} \label{CPII_is_linear}
The \enquote{Cherry Pickup II} problem has a linear solution.
\begin{proof}
Since count of cherries in cells are nonnegative values, then all values of $\g$ are nonnegative.
According to Lemma \ref{best_solve_in_lrtmspair}.1 and start positions of robots it is enough to find the sum of any $\lrtms(0,0,\W-1)$ pair in grid with nonnegative values.
According to definitions of $\lMaxes$ and $\rMaxes$ this sum is equal to $\lMaxes(0,\W-1)$ and $\rMaxes(0,0)$.
According to Lemma \ref{_Maxes_is_linear} we can find the tables $\lMaxes$ and $\rMaxes$ in $\O(\H\cdot\W)$.
\end{proof}
\end{Teo}

Algorithm implementation in Python showed in listings below.
Finding $\F$ showed in listing 1,
for $\lpath$ and $\rpath$ in listing 2,
for $\lMaxes$ and $\rMaxes$ in listing 3.
Main function with solution in listing 4.

\begin{Teo} \label{BtNxOut_is_linear}
If there are negative values in $\g$, but there are no bottlenecks, then problem can be solved by finding maximum sum of two node-disjoint paths on $\g$.
\begin{proof}
Since $\g$ don't has bootlenecks, then according to Lemma \ref{best_solve_in_lrtmspair}.2
and start positions of robots it is enough to find the sum of any $\lrtms(0,0,\W-1)$ pair.
According to definitions of $\lMaxes$ and $\rMaxes$ this sum is equal to $\lMaxes(0,\W-1)$ and $\rMaxes(0,0)$.
According to Lemma \ref{_Maxes_is_linear} we can find the tables $\lMaxes$ and $\rMaxes$ in $\O(\H\cdot\W)$.
\end{proof}
\end{Teo}

\subsection{Reducing of DM to finding maximum sum of two node-DP}

{\bf Problem description:}

Given a grid $\g_{DM}$ of size $N{\times}N$ with values in cells $0,1$ and $-1$:

$0$ means there is no diamond, but you can go through this cell;

$1$ means the diamond (i.e. you can go through this cell and pick up the diamond);

$-1$ means that you can't go through this cell.

We start at cell $(0,0)$ and reach the last cell $(N{-}1,N{-}1)$, and then return back to $(0,0)$ collecting maximum number of diamonds:

Going to last cell we can move only right and down;

Going back we can move only left and up.

\vspace{4px}
{\bf Solution:}

Let $\g_1$, $\g_2$ and $\g_3$ are grids of size $(2N{-}1){\times}(2N{-}1)$.
And $\g_4$ is grid of size $(3N{-}2){\times}(2N{-}1)$.
Denote $N-1$ as $n$.
%Denote maximum value og $\g_{DM}$ as M.
Then DM can be reduced to our LS of \CP\ (without proof of correctnes):
\begin{enumerate}
\vspace{-5px}
\item
Check matrix for reachability by 1 robot. If not richable then return 0.
\vspace{-5px}
\item\label{DM2CP2:2}
Turn matrix clockwise by $45^\circ$. I.e. for each $i=0,...,n, j=0,...,n$
$$
\g_1[i+j][n+i-j]=\g_{DM}[i][j].
$$
%\vspace{-5px}
\item\label{DM2CP2:3}
Add cells between horizontally neighboring cells.
Also add under upper cells, except $(0,n)$, by one cell.
Fill cell by $-10N$ if bottom neighbor is $-1$,
or both horizontally neighboring cells are $-1$.
Otherwise, fill by $0$.
I.e. for each $i=0,...,n, j=0,...,n$ where $i+j\ge1$
$$
\g_1[i{+}j{-}1][n{+}i{-}j]=\begin{cases}
-10N &\g_1[i{+}j][n{+}i{-}j]=-1,\\
-10N &\g_1[i{+}j{-}1][n{+}i{-}j{-}1]=-1\ and\ \g_1[i{+}j{-}1][n{+}i{-}j{+}1]=-1,\\
0 &otherwise.
\end{cases}
$$
%\vspace{-5px}
\item
Add corners, and fill them by $-10N$, except top and bottom rows.
Fill unvalued cells by $0$.
I.e. for each $i=0,...,2n, j=0,...,2n$
$$
\g_2[i][j]=\begin{cases}
-10N &0<i<2n\ and\ (i{+}j<n\ or\ i{+}j>3n\ or\ i{+}n<j\ or\ i>j{+}n),\\
0 &i=0\ and\ (j<n-1\ or\ j>n+1),\\
0 &i=2n\ and\ j\ne n,\\
\g_1[i][j] &otherwise.
\end{cases}
$$
%\vspace{-5px}
\item
Change values $-1$ by $-10N$.
I.e. for each $i=0,...,2n, j=0,...,2n$
$$
\g_3[i][j]=\begin{cases}
-10N &g_2[i][j]=-1,\\
\g_2[i][j] &otherwise.
\end{cases}
$$
%\vspace{-5px}
\item\label{DM2CP2:CP_BtNxOut}
Add on top the matrix of size $n{\times}(2n+1)$ filled by $0$.
I.e. for each $i=0,...,3n, j=0,...,2n$
$$
\g_4[i][j]=\begin{cases}
0 &i<n,\\
\g_3[i-n][j] &i\ge n.
\end{cases}
$$
%\vspace{-5px}
\item\label{DM2CP2:alg}
Apply  our LS of \CP\ for grid $\g_4$ and return answer.
\end{enumerate}

Since our algorithm looking for \paths\ without intersections,
therefore by instruction \ref{DM2CP2:3} we make double "road" with zero-sum for every reachable \path\ to avoid bottlenecks.
Therefore, after instruction \ref{DM2CP2:CP_BtNxOut}, due to Theorem
\ref{BtNxOut_is_linear}, we can get answer by applying our LS to $\g_4$.

First instruction can be checked by linear time using BFS.
Instructions \ref{DM2CP2:2} -- \ref{DM2CP2:CP_BtNxOut} are linear transformations.
And last instruction has linear comlexity.

More exatly this reducing used linear operations with values at most $\O(N^2)$.
I.e. these values have lengths $\O(\log(N))$ same as lengths of addresses to rows.
Therefore, we ignore these operations for complexity estimation.

\subsection{Some optimisation}

% first intersection (fi,fj)
\begin{Def}
Let $(fi,fj)$ is cell of first (least by rows) intersection of $\lpath$ with $\rpath$.
\end{Def}

% lPmax and rPmax
\begin{Def}
Let $\lPmax$ and $\rPmax$ are $\lrtms(0,0,\W-1)$ pair.
\end{Def}

% start from first intersection of ltrack and rtrack
\begin{Prop} \label{start_from_first_intersection}
Either $\lPmax[0,...,fi]=\lpath[0,...,fi]$ or
$\rPmax[0,...,fi]=\rpath[0,...,fi]$.
\begin {proof}
Suppose that one of these \tracks\ don't passes through intersection of $\lpath$ and $\rpath$, WLOG let it be $\rPmax$. Then $\rPmax$ don't intersect $\lpath$.
Then, due to Property \ref{between_intersections}.2, we get $\lPmax=\lpath$.
I.e. $\lPmax[0,...,fi]=\lpath[0,...,fi]$.

It remains to consider when $\lPmax$ intersect $\rpath$ in some $i_1$-th row and $\rPmax$ intersect $\lpath$ in some $i_2$-th row.
By Note \ref{l_RP} $fi\le\min\{i_1,i_2\}$.
WLOG let $i_1<i_2$, then due to Property \ref{between_intersections}.1 we get $\lPmax[0,...,i_1]=\lpath[0,...,i_1]$.
Since $fi\le i_1$, then $\lPmax[0,...,fi]=\lpath[0,...,fi]$.
\end {proof}
\end{Prop}

Using Lemma \ref{best_solve_in_lrtmspair} it is enough to find $\lrtms(0,0,W-1)$ pair
$\lPmax$ and $\rPmax$.
Also, due to Property \ref{start_from_first_intersection} either
$\lPmax(fi)=fj$ or $\rPmax(fi)=fj$.

WLOG let $\lPmax(fi)=fj$.
Let $maxPath(i,j)$ is path $p$ from $(0,\W-1)$ to $(i,j)$ with maximum sum.
Then, using Property \ref{start_from_first_intersection}, it is enough to find maximum of
$$\PS(\lpath[...,fi-1])+\PS(lp_j)+\PS(rp_j)+maxPath(fi,j)-g_{fi,j}$$
for each $j>fj$, where $lp_j$ and $rp_j$ are $\lrtms(fi,fj,j)$ pair.

Sum of $\lrtms(fi,fj,j)$ pair equal to $\lMaxes(fi,j)$.
For calculation of $maxPath(fi,j)$ for each $j>fj$ let consider next tables
% ltg table
\begin{Def}
Let $tg$ is grid:\\
$$
tg_{i,j} = \begin{cases}
-\infty &i\ge fi\ or\ i<j<\W-1-i,\\
\g_{i,j} &i < fi\ and\ (j\le i\ or\ j\ge\W-1-i).
\end{cases}
$$
\end{Def}

% udF(g) table
\begin{Def}
for $j=0,...\W{-}1$ the $udF_{i,j}(g')$ is table defined under grid $g'$ as:\\
$$
udF_{i,j}(g') = \begin{cases}
g'_{i,j} &i=0,\\
g'_{i,j} + \max\{udF_{i-1,j-1}(g'), udF_{i-1,j}(g'), udF_{i-1,j+1}(g')\} &i=1,...,\H-1.
\end{cases}
$$
\end{Def}

Similarly to $\F$ the $udF$ allows to find the \path\ with maximum sum.
For $j<fj$ the $udF_{fi,j}(tg)$ gives sum of \path\ with maximum sum between cells $(fi,j)$ and $(0,0)$.
And for $j>fj$ the $udF_{fi,j}(tg)$ gives maximum sum of \path\ between $(fi,j)$ and $(0,W-1)$.
Thus $maxPath(fi,j)=udF_{fi,j}(tg)$ for any $j>fj$.

Then, for solve our task we can find
$$lMax=\max_{j=fj+1,...,\W-1}\{\lMaxes(fi,j)+udF_{fi,j}(tg)-\g_{fi,j}\}+\F_{0,0}-\F_{fi-1,fj},$$
$$rMax=\max_{j=0,...,fj-1}\{\rMaxes(fi,j)+udF_{fi,j}(tg)-\g_{fi,j}\}+\F_{0,\W-1}-\F_{fi-1,fj}.$$
Then $\max\{lMax,rMax\}$ is required answer.

\subsection{Linear solutions for some extensions}
Let $0\le d_i<\W$ for each $i>0$.
Then rule (r1) can be extended as
\begin{itemize}
\vspace{-5px}
\item[(r1')]
From cell $(i-1,j)$ robots can move to cell $(i,j-d_i)$, $(i,j-d_i+1)$, ... or $(i,j+d_i)$.
\end{itemize}

Note that all Properties, Lemmas and Theorems can be generalized for extended rule (r1').
Therefore further we assume that it is true.

The length of input data is the length of grid plus the length of vector $d$.
Thus, the length of input data is $\Theta(\H\cdot\W)$.
Let prove that there are LS i.e. with complexity $\O(\H\cdot\W)$.

Let $SWM_{v, w}(j)=\max\{v(j-w), ..., v(j+w)\}$ where $v$ is vector.
$SWM_{v, w}$ is sliding window maximum (SWM) with window size $2w+1$.
The SWM is well known structure in programming, and can be defined as array of maximums of each subarray of size $2w+1$ in $v$.
SWM has $\langle\O(|v|), \O(1)\rangle$ complexity.
I.e. array $SWM_{v, w}$ can be prepared in $\O(|v|)$, and (after preparing) the value
$SWM_{v, w}(j)$ can be obtained in $\O(1)$ for each $j$ (as in ~\cite{bib:SWM}).
Then $\F$ can be extended as
$$
\F_{i,j} = \begin{cases}
0 &i=\H,\\
\g_{i,j} &i=\H-1,\\
\g_{i,j} + SWM_{R_{i+1, \F}, d_{i+1}}(j) &i=0,...,\H-2
\end{cases}
$$

where $R_{i,\F}$ is vector of length $\W{+}2d_i$ such that
$$
R_{i,\F}(j) = \begin{cases}
0 &-d_i\le j<0\ or\ \W\le j<\W+d_i,\\
\F_{i,j} &0\le j<\W.
\end{cases}
$$

Then each row for $\F, R$ and $SWM$ can be found sequentially: the first $\F_{H,*}$, then\\
$\F_{H{-}1,*}\rightarrow R_{H{-}1,\F}\rightarrow
SWM_{R_{H{-}1,\F}, d_{H{-}1}}\rightarrow\F_{H{-}2,*}\rightarrow
%R_{H{-}2,F}\rightarrow SWM_{d_{H-2}}(R_{H-2,\F},*)\cdots\\
\cdots\rightarrow R_{1,F}\rightarrow SWM_{R_{1,\F}, d_{1}}\rightarrow\F_{0,*}$.

Since $SWM_{R_{i,\F}, d_{i}}$ can be found in $\O(\W)$ for each $i$, then table $\F$
can be found in $\O(\H{\cdot}\W)$.

\vspace{4px}
Let prove that $\lMaxes$ and $\rMaxes$ can be found in $\O(\H{\cdot}\W)$.

Assume that $i, j, max_1, max_2$ and $max_3$ are designations from induction step of
Lemma \ref{_Maxes_is_linear}.

Let $b_1'=\max\{\lpath(i{-}1){-}d_i,0\}$ and $b_2'=\lpath(i){-}1$.

Let $b_1''=\max\{j{-}d_i,\rpath(i),\lpath(i){+}1\}$ and $b_2''=\min\{j{+}d_i,\W{-}1\}$.

And let $b_1=\max\{0,\lpath(i{-}1){-}d_i\}=b_1'$, $b_2=\lpath(i){-}1=b_2'$ and
$b_3=\max\{\rpath(i){+}1,j{-}d_i\}$, $b_4=\min\{j{+}d_i,\W{-}1\}$.

I.e. $b_1', b_2'$ are extended $b_1, b_2$ from case 1 of Lemma \ref{_Maxes_is_linear},
$b_1'', b_2''$ are extended $b_1, b_2$ from case 2, and
$b_1, b_2, b_3, b_4$ are extended $b_1, b_2, b_3, b_4$ from case 4.

$\max_1(j), max_2(j)$ and $max_3(j)$ can be found in $\O(1)$ using precalculated the SWM with window size $2d_i+1$ for $i$-th row of $\lMaxes, \rMaxes$ and $\F$.

Let $M_{li}$ is vector defined between positions $b_1''-d_i$ and $\W+d_i$ such that
$M_{li}(k)=\lMaxes(i,k)$ for each $b_1''\le k<\W$, and $M_{li}(k)=0$ for each $b_1''-d_i\le k<b_1''$ and $\W\le k\le\W+d_i$.
Then
$$
max_2(j)=\max_{k=b''_1,...,b''_2}\{\lMaxes(i,k)\}=SWM_{M_{li}, d_i}(j).
$$
Let $M_{ri}=\max_{k=b'_1,...,b'_2}\{\rMaxes(i,k)\}$ i.e. $max_1(j)=M_{ri}$ independ on $j$.

Let $F_i$ is vector defined between positions $b_3-d_i$ and $\W+d_i$ such that
$F_i(k)=\F(k)$ for each $b_3\le k<\W$, and $F_i(k)=0$ for each $b_3-d_i\le k<b_3$ and $\W\le k\le\W+d_i$.
Then
$$
max_3(j)=\max_{k=b_1,...,b_2}\{\rMaxes(i,k)\}+\max_{k=b_3,...,b_4}\{\F_{i,k}\}-\F_{i,\rpath(i)}
=M_{ri}+SWM_{F_i, d_i}(j)-\F_{i,\rpath(i)}.
$$

I.e. $max_1(j), max_2(j)$ and $max_3(j)$ can be found in $\O(1)$ with prepared $SWM_{M_{li}, d_i}, M_{ri}$ and $SWM_{F_i, d_i}$ for each $j$.

The $M_{ri}$ can be found in $\O(\W)$ and doesn't depend on $j$.
I.e. $M_{ri}$ can be represented as structure with $\langle\O(\W), \O(1)\rangle$ complexity.
The $SWM$ can be found for $M_{li}$ and $F_i$ with window $2d_i+1$ in
$\O(\W+2d_i)=\O(\W)$ for any row.
I.e. $SWM_{F_i, d_i}$ and $SWM_{M_{li}, d_i}$ are structures with
$\langle\O(\W), \O(1)\rangle$ complexity.

Thus every row of $\lMaxes$ and $\rMaxes$ can be found in $\O(\W)$.
I.e. this extension can be solved in $\O(\H\cdot\W)$ i.e. has linear solution.

\vspace{4px}
And another natural extension of \CP\ we formulate as
\begin{Con} \label{con1}
Let $n>0$ and $\W\ge n$.
And let there are $n$ robots located on different cells in the top row of $\g$, which moves by rules (r1), (r2) and (r3) to bottom row.
Then exists an algorithm for finding the maximum number of cherries, that can be collected by these robots, in $\O(\H\cdot\W\cdot2^n)$.
\end{Con}

For $n=1$ using $\F_{0,j}(\g)$ we get a proof of this Conjecture immediately for robot at $j$-th column.

For $n=2$ let robots starts from $j_1$ and $j_2$ columns where $j_1<j_2$.
Consider 2 cases:
\begin{enumerate}
\item
When $j_2-j_1>2\H$ then any \paths\ of robots don't intersect with each other.
Then this case can be reduced to sum of 2 independent solutions for $n=1$.
\item
$j_2-j_1\le2\H$ then all reachable columns by these robots in interval from $j_1-\H$ to $j_2+\H$.
Then we can get subgrid of size $\H{\times}(4\H)$ contains this interval
of all reachable columns.
Let denote this subgrid as $\g_d$.
Let $\g_u$ is grid of size $(2\H){\times}(4\H)$ with zeros.
Then let $\g'$ obtained by attaching the $\g_u$ under the $\g_d$.
Thus, we get $\g'$ of size $(3\H){\times}(4\H)$.

Now let $m$ is maximum value of $\g'$.
Then let $\g''$ is $\g'$ but with increased values by $m\cdot\H$ in cells $(2\H, j_1)$ and $(2\H, j_2)$.
Then after applying our LS for $\g''$ we get the sum of 2 DP, passes through the cells $(2\H, j_1)$ and $(2\H, j_2)$ with maximum sum M.
Then required value is $M-2m\cdot\H$.

Thus, we reduce the case $n=2$ to \CP\ by linear time.
Then using Theorem \ref{CPII_is_linear} we confirm our Conjecture for $n=2$.
\end{enumerate}

\newpage
\lstset{style=mystyle}

% F
\begin{lstlisting}[language=Python, caption=Search for maximums table for single robot (tested)]
import numpy as np

def get_F(g):
    H = len(g)
    W = len(g[0])
    F = np.empty((H, W))                 # create table HxW

    F[H-1] = g[H-1].copy()               # copy last row

    for i in reversed(np.arange(0, H-1)):#i = H-2, ..., 0
        F[i][0]   = g[i][0]   + max(F[i+1][0],   F[i+1][1])
        F[i][W-1] = g[i][W-1] + max(F[i+1][W-2], F[i+1][W-1])
        for j in np.arange(1, W-1):      # j = 1, ..., W-2
            F[i][j] = g[i][j] + max(F[i+1][j-1], F[i+1][j], F[i+1][j+1])

    return F
\end{lstlisting}

% lpath, rpath
\begin{lstlisting}[language=Python, caption=Search for $\lpath$ and $\rpath$ (tested)]
def get_bounds(F):
    H = len(F)
    W = len(F[0])
    lp = np.arange(0, H)# lp = [0, ..., H-1]
    rp = np.arange(0, H)# rp = [0, ..., H-1]

    lp[0] = 0
    rp[0] = W - 1

    for i in np.arange(1, H):   # i = 1, ..., H-1
        lj = lp[i] = lp[i-1]
        if lj > 0 and F[i][lj-1] >= F[i][lj]:
            lp[i] = lj - 1
        if lj < W-1 and F[i][lp[i]] < F[i][lj+1]:
            lp[i] = lj + 1

        rj = rp[i] = rp[i-1]
        if rj < W-1 and F[i][rj+1] >= F[i][rj]:
            rp[i] = rj + 1
        if rj > 0 and F[i][rp[i]] < F[i][rj-1]:
            rp[i] = rj - 1

    return lp, rp
\end{lstlisting}

% *Maxes shorter
\begin{lstlisting}[language=Python, caption=Search for $\lMaxes$ and $\rMaxes$ tables (short version. tested)]
def get_max(fromk, tok, Table, i):
    _max = float('-inf')
    for k in np.arange(fromk, tok+1):# k = fromk,...,tok
        _max = max(_max, Table[i][k])
    return _max

def get_M(g, F, lp, rp):
    H, W = len(F), len(F[0])
    Ml, Mr = np.empty((H, W)), np.empty((H, W))

    # base case M*[H-1]
    lj = max(rp[H-1], lp[H-1]+1)
    for j in np.arange(lj, W):  # j = max(rp[H-1], lp[H-1]+1),...,W-1
        Ml[H-1][j] = g[H-1][lp[H-1]] + g[H-1][j]

    rj = min(lp[H-1], rp[H-1]-1)
    for j in np.arange(0, rj+1):# j = 0,...,min(lp[H-1], rp[H-1]-1)
        Mr[H-1][j] = g[H-1][rp[H-1]] + g[H-1][j]

    # induction step M*[0,...,H-2]
    for i in reversed(np.arange(0, H-1)): # i = H-2,...,0
        Mri = get_max(max(0, lp[i]-1), lp[i+1]-1, Mr, i+1)
        Mli = get_max(rp[i+1]+1, min(W-1, rp[i]+1), Ml, i+1)

        # Ml[i] search
        for j in np.arange(max(lp[i]+1,rp[i]), W):
            max1, max2 = 0, 0

            # case lPmax(i+1)<lp(i+1)
            if max(lp[i],1) <= lp[i+1]:
                max1 = get_max(max(rp[i+1], j-1), min(j+1, W-1),
                               F, i+1) + Mri - F[i+1][rp[i+1]]

            # case lPmax(i+1)=lp(i+1)
            if lp[i+1]+2 <= W:
                max2 = get_max(max(j-1, rp[i+1], lp[i+1]+1),
                               min(j+1, W-1),
                               Ml, i+1)

            Ml[i][j] = g[i][lp[i]] + g[i][j] + max(max1,max2)

        # Mr[i] search
        for j in np.arange(0, min(lp[i],rp[i]-1)+1):
            max1, max2 = 0, 0

            # case rPmax(i+1)>rp(i+1)
            if rp[i+1] <= min(W-2,rp[i]):
                max1 = get_max(max(0, j-1),  min(j+1, lp[i+1]),
                               F, i+1) + Mli - F[i+1][lp[i+1]]

            # case rPmax(i+1)=rp(i+1)
            if 1 <= rp[i+1]:
                max2 = get_max(max(j-1, 0),
                               min(j+1, lp[i+1], rp[i+1]-1),
                               Mr, i+1)

            Mr[i][j] = g[i][rp[i]] + g[i][j] + max(max1,max2)

    return Ml, Mr
\end{lstlisting}

% main alg.
\begin{lstlisting}[language=Python, caption=Main algorithm (tested)]
def Pickup_Cherries_II(grid):
    W      = len(grid[0])
    F      = get_F(grid)
    lp, rp = get_bounds(F)
    Ml, Mr = get_M(grid, F, lp, rp)

    return Ml[0][W-1]
\end{lstlisting}

\end{document}